\documentclass[aps,floats]{revtex4-2}
\usepackage{amsmath,amssymb}
\usepackage{graphicx,epsfig}
\usepackage[greek,english]{babel}
\usepackage{bbold}

\begin{document}
\bibliographystyle {plain}

\pdfoutput=1
\def\oppropto{\mathop{\propto}} 
\def\opsimeq{\mathop{\simeq}}
\def\opoverderline{\mathop{\overline}}
\def\operarrow{\mathop{\longrightarrow}}
\def\opsim{\mathop{\sim}}

\def\opmin{\mathop{\min}} 
\def\opmax{\mathop{\max}} 
\def\oplim{\mathop{\lim}}

\title{ Supersymmetric properties of one-dimensional reversible Markov generators  \\
with the links to Markov-dualities and to shape-invariance-exact-solvability }


\author{C\'ecile Monthus}
\affiliation{Universit\'e Paris-Saclay, CNRS, CEA, Institut de Physique Th\'eorique, 91191 Gif-sur-Yvette, France}


\begin{abstract}
For one-dimensional reversible diffusion process involving the force $F(x)$ and the diffusion coefficient $D(x)$, the continuity equation $\partial_t P_t(x)=- \frac{\partial}{\partial x}J_t(x)$ gives the dynamics of the probability $P_t( x)$ in terms of the divergence of the current $J_t( x)=F(x)P_t(x)-D(x)\frac{\partial}{\partial x} P_t(x)={\cal J}P_t( x)$ obtained from $P_t( x) $ via the application of the first-order differential current-operator ${\cal J}$. So the dynamics of the probability $P_t( x)$ is governed by the factorized Fokker-Planck generator ${\cal F}=-\frac{\partial}{\partial x}{\cal J}$, while the dynamics of the current $J_t( x)$ is governed by its supersymmetric partner ${\hat {\cal F} }= - {\cal J}\frac{\partial}{\partial x}$, so that their right and left eigenvectors are directly related using the two intertwining relations ${\cal J}{\cal F}=-{\cal J}\frac{\partial}{\partial x}{\cal J}={\hat {\cal F}}{\cal J}$ and ${\cal F}\frac{\partial}{\partial x}=-\frac{\partial}{\partial x}{\cal J}\frac{\partial}{\partial x}=\frac{\partial}{\partial x}{\hat {\cal F} }$. We also describe the links with the supersymmetric quantum hermitian Hamiltonian $H=Q^{\dagger} Q$ that can be obtained from the Fokker-Planck generator ${\cal F}$ via a similarity transformation, and with the factorization of the adjoint $ {\cal F}^{\dagger}=\frac{d}{dm(x)}\frac{d}{ds(x)} $ of the mathematical literature in terms of the scale function $s(x)$ and the speed measure $m(x)$. We then analyze how the supersymmetric partner ${\hat {\cal F} } = - {\cal J}  \frac{\partial}{\partial x}$ can be re-interpreted in two ways: (1) as the adjoint ${\mathring {\cal F}}^{\dagger} ={\mathring {\cal J} }^{\dagger} \frac{\partial}{\partial x}$ of  the Fokker-Planck generator ${\mathring {\cal F}}=- \frac{\partial}{\partial x}{\mathring {\cal J} }$ associated to the dual force ${\mathring F}(x)=-F(x)-D'(x)$, that reformulates various known Markov dualities at the level of the operator identity ${\hat {\cal F} }={\mathring {\cal F}}^{\dagger} $; (2)  as the non-conserved Fokker-Planck generator ${\tilde {\cal F}}_{nc} =  -\frac{\partial}{\partial x}{\tilde {\cal J}}-{\tilde K }(x)$  involving the force ${\tilde F}(x)=F(x) +D'(x)$ and the killing rate ${\tilde K }(x)=-F'(x)-D''(x)$, that explains why Pearson diffusions involving a linear force $F(x)$ and a quadratic diffusion coefficient $D(x)$ are exactly solvable for their spectral properties via the shape-invariance property involving constant killing rates. Finally, we describe how all these ideas can be also applied to Markov jump processes on the one-dimensional lattice with arbitrary nearest-neighbors transition rates $w(x \pm 1,x)$, also called Birth-Death processes.

\end{abstract}

\maketitle

\section{ Introduction }

\subsection{ On the notion of supersymmetry for reversible and irreversible Markov generators }

In the field of time-homogeneous stationary Markov models, it is essential to distinguish
between reversible processes satisfying detailed-balance and irreversible processes breaking detailed-balance.
In particular, their generators have very different properties as we now recall.

One of the most well-known property of reversible Markov generators 
that is explained in textbooks \cite{gardiner,vankampen,risken} 
and that has been extensively used in many specific applications to various models (see for instance \cite{glauber,Felderhof,siggia,referee,kimball,peschel,jpb_antoine,pierre,texier,us_eigenvaluemethod,Castelnovo,c_lyapunov,us_gyrator,us_kemeny,c_pearson,c_boundarydriven,c_largedevDBsusy}), is that 
they can be transformed into hermitian quantum Hamiltonians
via similarity transformations. This property yields that the relaxation towards equilibrium involves only real eigenvalues. In addition, the corresponding quantum Hamiltonians are not arbitrary but can be factorized into the supersymmetric form $H=Q^{\dagger} Q$ in terms of another operator $Q$ and its adjoint $Q^{\dagger}$ (see the reviews \cite{review_susyquantum,Mielnik_review,sasaki_reviewSusyQM,Review_factorization}
with various scopes on supersymmetric quantum mechanics in continuous space).
This supersymmetric factorization is very standard for diffusion processes (see for instance  \cite{jpb_antoine,pierre,texier,c_lyapunov,us_gyrator,us_kemeny,c_pearson,c_boundarydriven}) 
but is also very useful for Markov jump processes \cite{c_boundarydriven,c_largedevDBsusy}.

For irreversible Markov processes involving steady currents, 
this correspondence towards supersymmetric hermitian quantum Hamiltonians $H=Q^{\dagger} Q$
is lost, so that eigenvalues can be complex and can produce oscillations.
However, as discussed in detail in the recent work \cite{c_susySVD}, the continuity equation 
formulated either in continuous space or in discrete space yields that the Markov generator
 is always naturally factorized into the product $ {\bold {div}} . {\bold J} $  of the divergence 
operator ${\bold {div}}$ and of the current operator ${\bold J}$,
while its supersymmetric partner ${\bold J}   {\bold {div}} $, where the two operators appear in the opposite order, governs the dynamics of the currents. It is important to stress that the probabilities and the currents 
a priori live in very different spaces (see \cite{c_susySVD} for more details) : 
(i) for Fokker-Planck generators in dimension $d>1$, the probability $P_t(\vec x)$ is a scalar function, while the current $\vec J_t(\vec x)$ is a d-dimensional vector;
(ii) for Markov jump processes, the probability $P_t(x)$ lives on the $N$ discrete configurations $x$, while the currents lives on the $M= (N-1)+C$ links between configurations, where $C=M-(N-1)\geq 1$ is the number of independent cycles that may carry steady currents. In both cases, the configuration space for the currents $J_t(x)$ is thus bigger than the configuration space for the probability $P_t(x)$. 

However in dimension $d=1$, the situation is completely different 
since both the probability $P_t(x)$ and the current $J_t(x)$ live in one dimension, 
while a non-vanishing steady current is possible only for periodic boundary conditions 
or for boundary-driven models where the boundary probabilities are fixed by external reservoirs.
In both cases, as discussed in detail in the respective recent works \cite{c_lyapunov} and \cite{c_boundarydriven},
it is nevertheless still possible and very useful to continue to use the mapping towards supersymmetric quantum Hamiltonians $H=Q^{\dagger} Q$ in the bulk, even if the right and left eigenvectors will be nevertheless different as a consequence of the imposed boundary conditions that break detailed-balance and produce a non-vanishing steady current.


\subsection{ Goals and organization of the present paper concerning one-dimensional reversible Markov processes }

For one-dimensional reversible Markov processes, 
the standard mapping towards hermitian quantum Hamiltonian $H=Q^{\dagger} Q$ has been extensively used as recalled above, but the supersymmetric-partner hermitian quantum Hamiltonian $Q Q^{\dagger} $ 
is then introduced as a technical trick without any direct physical meaning for the initial Markov process.
In the present paper, we thus wish to consider instead the probabilities and the currents on the same footing,
with the corresponding natural factorization $ {\bold {div}} . {\bold J} $ of the generator coming from the continuity 
equation as described above with the following advantages :

(a) The supersymmetric-partner ${\bold J}   {\bold {div}} $ has then a very clear physical relevance
since it governs the dynamics of the current.

(b) This perspective is also useful to make the link with the mathematical literature
where it is standard to factorize the adjoint of Markov generator ${\bold J}^{\dagger}   {\bold {div}}^{\dagger} $  that governs the dynamics of observables
  in terms of the scale function $s(x)$ and the speed measure $m(x)$.
  
  After this physically-motivated unifying reformulation 
  of various known properties from the physical and mathematical literature,
  we will turn to the more novel part of the present work
where the goal is to analyze the properties of the 
supersymmetric partner ${\bold J}   {\bold {div}} $ governing the dynamics of the current :

(i) we will first analyze its spectral properties via the two intertwining relations
between the supersymmetric partners $ {\bold {div}} . {\bold J} $ and ${\bold J}   {\bold {div}} $.

(ii) we will then discuss two possible re-interpretations of the supersymmetric partner ${\bold J}   {\bold {div}} $,
namely as the adjoint generator of some dual-process conserving probabilities
 and alternatively as a non-conserved Markov generator involving some killing rates.

While all these ideas can be applied to Markov processes defined either in continuous space or discrete space,
there are important technical differences, so for clarity 
we have chosen to focus on diffusion processes in the main text
and to focus on Markov jump processes in the Appendices, with the following organization.

\subsubsection{ Organization of the main text concerning diffusion processes involving arbitrary force $F(x)$ and diffusion coefficient $D(x)$ }

 The main text is devoted to reversible Fokker-Planck generators in dimension $d=1$
 involving an arbitrary force $F(x)$ and an arbitrary diffusion coefficient $D(x)$
with the following organization:

$\bullet $ In section \ref{sec_diffusionPJ}, we stress that in diffusion processes, it is both technically convenient 
and physically insightful to consider the probability $P_t(x)$ 
and the current $J_t(x)$ on the same footing, and to describe their couplings
via two first-order differential operators.

$\bullet $ In section \ref{sec_diffusionP}, we describe how the dynamics of the
probability $P_t(x)$ alone is governed by the Fokker-Planck generator 
${\cal F}    \equiv  -  \frac{\partial}{\partial x} {\cal J} $ factorized into 
two first-order differential operators, and we recall many important properties.

$\bullet $ In section \ref{sec_diffusionJ}, we describe how the dynamics of
current $J_t(x)$ alone is governed by the supersymmetric partner 
${\hat {\cal F} }   \equiv  - {\cal J}  \frac{\partial}{\partial x}$ of 
the Fokker-Planck generator ${\cal F}  =  -  \frac{\partial}{\partial x} {\cal J}  $,
and we analyze its spectral properties using the corresponding intertwining relations ${\cal J}{\cal F}   =  - {\cal J} \frac{\partial}{\partial x} {\cal J} =  {\hat {\cal F} } {\cal J}$ and ${\cal F} \frac{\partial}{\partial x}  =  -  \frac{\partial}{\partial x} {\cal J} \frac{\partial}{\partial x} = \frac{\partial}{\partial x}{\hat {\cal F} }  $.

$\bullet $ In section \ref{sec_duality}, we explain how the supersymmetric partner ${\hat {\cal F} } = - {\cal J}  \frac{\partial}{\partial x}$ can be re-interpreted as the adjoint ${\mathring {\cal F}}^{\dagger} ={\mathring {\cal J} }^{\dagger} \frac{\partial}{\partial x}  $ of  the Fokker-Planck generator ${\mathring {\cal F}}=-  \frac{\partial}{\partial x}  {\mathring {\cal J} }$ associated to the dual force ${\mathring F}(x) =- F(x) -  D'(x)$, in order to reformulate various known Markov dualities at the level of the operator identity ${\hat {\cal F} } ={\mathring {\cal F}}^{\dagger} $.

$\bullet $ In section \ref{sec_nckilling}, we study how the supersymmetric partner ${\hat {\cal F} } = - {\cal J}  \frac{\partial}{\partial x}$ can be re-interpreted as the non-conserved Fokker-Planck generator ${\tilde {\cal F}}_{nc} =  -\frac{\partial}{\partial x} {\tilde {\cal J}} - {\tilde K }(x)$  involving the force $  {\tilde F }(x) =F(x) +D'(x)$ and the killing rate ${\tilde K }(x)=-F'(x)-D''(x)  $, and we explain why Pearson diffusions involving a linear force $F(x)$ and a quadratic diffusion coefficient $D(x)$ are exactly solvable for their spectral properties via the shape-invariance property involving constant killing rates.

Our conclusions are summarized in section \ref{sec_conclusion}.

\subsubsection{ Organization of the Appendices concerning Markov jump processes involving arbitrary transition rates $w(x \pm 1,x)$ }

 The various Appendices describe how each section of the main text can be adapted
to analyze instead reversible Markov jump processes on the one-dimensional lattice with arbitrary nearest-neighbors transition rates $w(x \pm 1,x)$, also called Birth-Death processes.
The presentation is shorter since the ideas are the same, but we stress some important technical differences.


\section{ Diffusion processes in terms of the probability $P_t(x)$ and the current $J_t(x)$}

\label{sec_diffusionPJ}

In this section, we introduce the notations that will be useful for all sections of the main text.
The main idea is that both technically convenient 
and physically insightful to consider the probability $P_t(x)$ 
and the current $J_t(x)$ on the same footing, and to describe their coupling
via two first-order differential operators.
This is somewhat similar to classical mechanics where it is well-known that 
the Hamiltonian formulation in terms of two first-order differentials equations for the position and the momentum is much more powerful than the Newton second-order equation for the position alone.

\subsection{ Probability $P_t(x)$ and current $J_t(x)$ coupled via two first-order differential operators}

The continuity equation for the probability $ P_t( x ) $ to be at position $x$ at time $t$
\begin{eqnarray}
 \partial_t P_t( x )    =  - \frac{\partial}{\partial x} J_t(x)
\label{continuity}
\end{eqnarray}
involves the divergence of the current 
\begin{eqnarray}
 J_t( x )  \equiv F(x)  P_t(x)   - D(x) \frac{\partial}{\partial x}  P_t(x) \equiv {\cal J} P_t(x)
\label{fokkerplanckcurrent}
\end{eqnarray}
that can be obtained from the probability $P_t(x)$ via the application of the first-order differential current-operator
\begin{eqnarray}
 {\cal J} \equiv F(x)   - D(x) \frac{\partial}{\partial x}  
\label{defJop}
\end{eqnarray}
that contains the force $F(x) $ and the diffusion coefficient $D(x)$.
Note that we will use the terminology 'force F(x)' as in many physical papers,
while another part of the literature instead prefers the word 'drift' with various other notations.


\subsection{ Vanishing-current Boundary Conditions $J_t(x_L) =0 = J_t(x_R)$ }

This dynamics of Eq. \ref{continuity}
on the interval $x \in ]x_L,x_R[$ has to be supplemented by boundary conditions 
at the left boundary $x \to x_L$ (that can be either finite or infinite $x_L=-\infty$)
and at the right-boundary $x \to x_R$ (that can be either finite or infinite $x_R=+\infty$).
In particular, the dynamics of the total probability $\left[ \int_{x_L}^{x_R} dx P_t( x ) \right] $
\begin{eqnarray}
 \partial_t \left[ \int_{x_L}^{x_R} dx P_t( x ) \right]   = \int_{x_L}^{x_R} dx \partial_t P_t( x )
 =  - \int_{x_L}^{x_R} dx \frac{\partial}{\partial x} J_t(x) = J_t(x_L) - J_t(x_R)
\label{conservTotal}
\end{eqnarray}
involves the difference of the currents $J_t(x_L) $ and $J_t(x_R) $ at the two boundaries.

In the present paper, we will focus only on the case of
\begin{eqnarray}
 \text{ Vanishing-current Boundary Conditions : } \ \ \ J_t(x_L) =0 = J_t(x_R) 
\label{vanishinhJ}
\end{eqnarray}
that can lead to a normalizable equilibrium steady state $P_{eq}(x)$ associated to vanishing steady current $J_{eq}=0$ (detailed-balance). Note that the terminology 'vanishing-current B.C.' is useful to describe together
the case of 'reflecting boundary conditions' 
when the diffusive particle is really able to reach the boundaries at the finite positions $x_L$ and $x_R$,
and the cases when the boundaries $x_L$ and  $x_R$ are infinite, or finite but cannot be really reached by the diffusive particle as a consequence of the specific forms of the force $F(x)$ and of the diffusion coefficient $D(x)$.

It is very important to stress the 
difference with two other boundary conditions that can lead to non-equilibrium steady states 
$P_{neq}(x)$ associated to a non-vanishing steady current $J_{neq} \ne 0$ 
that will not be discussed here :

$\bullet$ The case of Periodic Boundary Conditions on a ring $x_L \equiv x_R$ 
\begin{eqnarray}
 \text{ Periodic B.C. on a ring : } \ \ \ x_L \equiv x_R
\label{PeriodicBC}
\end{eqnarray}
has been analyzed from the supersymmetric point of view in \cite{c_lyapunov}.

$\bullet$  The case of Boundary-driven Boundary Conditions
where the two boundary values $P_t(x_L) =p_L$ and $P_t(x_R)=p_R$ are fixed by external reservoirs 
\begin{eqnarray}
 \text{ Boundary-driven B.C. : $P_t(x_L) =p_L$ and $P_t(x_R)=p_R$ fixed by external reservoirs }
\label{drivenBC}
\end{eqnarray}
has been analyzed from the supersymmetric point of view in \cite{c_boundarydriven}.


\subsection{ Observables involving the probability $P_t(x)$ and 
observables involving the current $J_t(x)$  }

The average of an observable $O(x)$ computed with the probability $P_t(x)$
\begin{eqnarray}
  \int_{x_L}^{x_R} dx O(x) P_t(x) = \langle O \vert P_t \rangle
 \label{Oav}
\end{eqnarray}
can be considered as the scalar product between the observable-bra $\langle O \vert $ and the probability-ket $\vert P_t \rangle$.

Its dynamics
can be analyzed using 
the continuity Eq. \ref{continuity} for $P_t(x)$ 
and performing one integration by parts using the vanishing-current Boundary Conditions of Eq. \ref{vanishinhJ}
\begin{eqnarray}
\partial_t  \bigg(  \int_{x_L}^{x_R} dx O(x) P_t(x)  \bigg)  && = \int_{x_L}^{x_R} dx O(x) \partial_t P_t(x)
= - \int_{x_L}^{x_R} dx O(x) \frac{\partial}{\partial x} J_t(x)  
\nonumber \\
&&=\int_{x_L}^{x_R} dx J_t(x) \frac{d O}{d x} - \bigg[ O(x) J_t(x)  \bigg]_{x_L}^{x_R}
\nonumber \\
&& =\int_{x_L}^{x_R} dx J_t(x) \frac{d O}{d x} \equiv  \langle \frac{d O}{d x} \vert J_t \rangle 
 \label{Oavdyn}
\end{eqnarray}

\subsection{ Discussion } 

In this section, we have considered the probability $P_t(x)$ and the current $J_t(x)$ on the same footing
and we have written their coupling via two first-order differential operators.
In order to obtain a closed dynamical equation for each of them,
one can either eliminate the current $J_t(x)$ to obtain the standard Fokker-Planck dynamics for the probability $P_t(x)$ alone, as recalled in the next section \ref{sec_diffusionP},
or one can eliminate the probability $P_t(x)$ to obtain the dynamics for the current $J_t(x)$ alone,
as described in the further section \ref{sec_diffusionJ}.


\section{ Factorized generator ${\cal F}    \equiv  -  \frac{\partial}{\partial x} {\cal J}$ governing the dynamics of the probability $P_t(x)$  }

\label{sec_diffusionP}

In this section, we describe how the dynamics of the
probability $P_t(x)$ is governed by the second-order differential Fokker-Planck generator 
${\cal F}    \equiv  -  \frac{\partial}{\partial x} {\cal J} $ factorized into 
the two first-order differential operators $\frac{\partial}{\partial x} $ and $ {\cal J} $.
 We recall various important consequences.


\subsection{ Fokker-Planck generator 
${\cal F}   =  -  \frac{\partial}{\partial x} {\cal J} $ factorized into two first-order differential operators}

Putting Eqs \ref{continuity} and \ref{fokkerplanckcurrent} together
yields that the dynamics of the probability $P_t(x)$ 
\begin{eqnarray}
 \partial_t P_t( x )    =  -  \frac{\partial}{\partial x} {\cal J} P_t(x) \equiv {\cal F} P_t(x)
\label{fokkerplanck}
\end{eqnarray}
is governed by the second-order differential Fokker-Planck generator
that is naturally factorized into the product of the two first-order differential operators 
\begin{eqnarray}
{\cal F}  &&  \equiv  -  \frac{\partial}{\partial x} {\cal J} =-\frac{\partial}{\partial x} \left(  F(x)   - D(x) \frac{\partial}{\partial x}  \right)
\label{fokkerplanckgeneratorfactorized}
\end{eqnarray}
while its expanded version reads
\begin{eqnarray}
{\cal F}  &&     =-\frac{\partial}{\partial x}   F(x)   + \frac{\partial}{\partial x}  D(x) \frac{\partial}{\partial x}
\nonumber \\
&& = - F'(x) + \left[ D'(x) -  F(x) \right] \frac{\partial}{\partial x} +   D(x) \frac{\partial^2}{\partial x^2}  
\label{fokkerplanckgeneratorexpanded}
\end{eqnarray}

It is useful to replace the force $F(x)$ by the potential $U(x)$ with its derivative $U'(x)$
\begin{eqnarray}
U'(x) && = - \frac{F(x)}{D(x)}
\nonumber \\
U(x) && \equiv - \int_{x_0}^x dy \frac{F(y)}{D(y)} \ \ \text{ with some convenient $x_0$ }
 \label{Ux}
\end{eqnarray}
in order to rewrite the current operator of Eq. \ref{defJop}
in the factorized form
\begin{eqnarray}
 {\cal J} && = -D(x) U'(x)   - D(x) \frac{\partial}{\partial x}  
   = - D(x) \left[ U'(x) + \frac{\partial}{\partial x} \right] 
   = - D(x) e^{- U(x) }  \frac{\partial}{\partial x}  e^{ U(x) }  
\label{defJopU}
\end{eqnarray}
so that the Fokker-Planck generator of Eq. \ref{fokkerplanckgeneratorfactorized}
becomes
\begin{eqnarray}
 {\cal F}   =  -  \frac{\partial}{\partial x} {\cal J} 
=\frac{\partial}{\partial x}  D(x) e^{- U(x) }  \frac{\partial}{\partial x}  e^{ U(x) }  
\label{fokkerplanckgeneratorfactorizedU}
\end{eqnarray}
This fully factorized form of the Fokker-Planck generator is very suggestive since it involves
 the standard Boltzmann factor $e^{- U(x) } $ \cite{zwanzig,ceccato}
and is useful to write the explicit solutions
of the second-order differential equations $0 = {\cal F} P_*(x) $ and $0= {\cal F}^{\dagger} O (x) $ 
as described in the next subsections.


\subsection{ Analyzing the steady-state $ P_*(x)$
from the two independent explicit solutions of 
$0 = {\cal F} P_*(x)  =  -  \frac{\partial}{\partial x} {\cal J} P_* (x) $ }

An important property is whether the Fokker-Planck dynamics of Eq. \ref{fokkerplanck}
leads to a steady-state $P_*(x)$, so one needs to consider 
the second-order differential equation involving the factorized Fokker-Planck generator ${\cal F} $ of Eq. \ref{fokkerplanckgeneratorfactorizedU}
\begin{eqnarray}
0= {\cal F} P_*(x)  =  -  \frac{\partial}{\partial x} {\cal J} P_*(x)
=\frac{d}{d x} \left(  D(x) e^{- U(x) }  \frac{d}{d x}  \left[ e^{ U(x) }  P_*(x) \right] \right)
\label{fokkerplanckgeneratorfactorizedUzero}
\end{eqnarray}
The first integration yields that the steady current associated to $P_*(x)$
should be constant ${\cal J} P_*(x) =J_* $ 
\begin{eqnarray}
 J_* =  {\cal J} P_*(x) = -  D(x) e^{- U(x) }  \frac{d}{d x}  \left[ e^{ U(x) }  P_*(x) \right]  
\label{jsteady}
\end{eqnarray}
The second integration leads to the linear combination 
\begin{eqnarray}
  P_*(x) = c_{eq} \rho_{eq}(x) - J_* \rho_{neq}(x)
\label{CLpsat}
\end{eqnarray}
of two independent solutions, where $\rho_{eq}(x) $ is the equilibrium solution satisfying detailed-balance
written as the standard Boltzmann factor $e^{- U(x) } $
\begin{eqnarray}
\rho_{eq}(x) && = e^{- U(x)} \ \ \text{ with vanishing steady current (detailed-balance)}:  \ \ \ J_{eq}={\cal J} \rho_{eq}(x) =0
\label{rhoeq}
\end{eqnarray}
that will be selected by the vanishing-current Boundary Conditions of Eq. \ref{vanishinhJ} considered in the present paper, while $ \rho_{neq}(x)$ is a non-equilibrium solution 
\begin{eqnarray}
\rho_{neq}(x) && = e^{- U(x)} \int_0^x dy \frac{ e^{U(y) }  } { D(y) } \ \ \ 
\text{ with non-vanishing steady current }:  \ \ \ {\cal J} \rho_{neq}(x) = -1
\label{rhoneq}
\end{eqnarray}
that is relevant for the periodic B.C. of Eq. \ref{PeriodicBC} (see \cite{c_lyapunov} and references therein)
or for boundary-driven B.C. involving external reservoirs of Eq. \ref{drivenBC} 
(see \cite{c_boundarydriven} and references therein).


\subsection{ Factorized adjoint operator 
${\cal F}^{\dagger}  =  {\cal J}^{\dagger}  \frac{\partial}{\partial x}  $ governing the bulk dynamics of observables $O(x)$ }

For the observable of Eq. \ref{Oav},
the dynamics of Eq. \ref{Oavdyn}
can be rewritten using the current $J_t(x)$ of Eq. \ref{fokkerplanckcurrent}
and making an integration by parts that will introduce boundary terms
\begin{eqnarray}
\langle \frac{d O}{d x} \vert J_t \rangle && =\int_{x_L}^{x_R} dx \frac{d O (x)}{d x}  J_t(x) 
= \int_{x_L}^{x_R} dx \frac{d O(x)}{d x} \bigg(  F(x)   P_t(x) - D (x)   \frac{\partial}{\partial x} P_t(x) \bigg)
\nonumber \\
&& =\int_{x_L}^{x_R} dx    P_t(x) \bigg(  F(x)  \frac{d O(x)}{d x} +  \frac{\partial}{\partial x} \left[ D(x) \frac{d O(x)}{d x} \right] \bigg)
 - \bigg[ P_t(x)  D(x) O'(x) \bigg]_{x_L}^{x_R}
 \nonumber \\
&& =\int_{x_L}^{x_R} dx    P_t(x)  \bigg( {\cal F}^{\dagger}  O(x) \bigg)
 - \bigg[ P_t(x)  D(x) O'(x) \bigg]_{x_L}^{x_R}
 \equiv \langle {\cal F}^{\dagger} O \vert P_t \rangle  - \bigg[ P_t(x)  D(x) O'(x) \bigg]_{x_L}^{x_R}
 \label{Oavdyn2emeinteg}
\end{eqnarray}
while the bulk contribution is given by the scalar product $\langle {\cal F}^{\dagger} O \vert P_t \rangle $
that involves the adjoint ${\cal F}^{\dagger} $ of the Fokker-Planck generator
$ {\cal F}$
\begin{eqnarray}
  {\cal F}^{\dagger}   =  {\cal J}^{\dagger}  \frac{\partial}{\partial x}
  && = \left(F(x)   + \frac{\partial}{\partial x}   D(x) \right) \frac{\partial}{\partial x}
\nonumber \\
&&  = \left[F(x)+D'(x)\right]\frac{\partial}{\partial x}   + D(x) \frac{\partial^2}{\partial x^2}  
  \label{FPgeneratorprodadjoint}
\end{eqnarray}

Its factorization into the product of the two first-order differential operators $ \frac{\partial}{\partial x}$
and the adjoint ${\cal J}^{\dagger} $ of the current operator of Eq. \ref{defJop} and \ref{defJopUadjoint}
\begin{eqnarray}
 {\cal J}^{\dagger} && = F(x)   + \frac{\partial}{\partial x}   D(x) 
 = \left[F(x)+D'(x)\right] + D(x) \frac{\partial}{\partial x}  
\nonumber \\
 && =e^{ U(x) }   \frac{\partial}{\partial x} D(x) e^{- U(x) } 
 \label{defJopUadjoint}
\end{eqnarray}
is useful to analyze whether some observables $O(x)$ are conserved by the full dynamics
of Eqs \ref{Oavdyn}
and \ref{Oavdyn2emeinteg}
\begin{eqnarray}
\partial_t  \bigg(  \int_{x_L}^{x_R} dx O(x) P_t(x)  \bigg)   =
  \langle \frac{d O}{d x} \vert J_t \rangle
  = \langle {\cal F}^{\dagger} O \vert P_t \rangle  - \bigg[  P_t(x)  D(x) O'(x)  \bigg]_{x_L}^{x_R}
 \label{Oavdynfull}
\end{eqnarray}
as described in the next subsection.


\subsection{ Analyzing conserved observables $O(x)$ from the two independent solutions of 
$0= {\cal F}^{\dagger} O (x) =  {\cal J}^{\dagger}  \frac{\partial}{\partial x} O(x)$}

\label{subsec_conservO}

An observable $O(x)$ will be conserved by the bulk-dynamics of Eq. \ref{Oavdyn}
if it satisfies the second-order differential equation involving the factorized adjoint ${\cal F}^{\dagger} $ of Eq. \ref{FPgeneratorprodadjoint}
\begin{eqnarray}
0= {\cal F}^{\dagger} O (x) =  {\cal J}^{\dagger}  \frac{\partial}{\partial x} O(x)
=e^{ U(x) }   \frac{d}{d x} \left[ D(x) e^{- U(x) }  \frac{d O(x)}{d x} \right]
\label{adjointzero}
\end{eqnarray}
The first integration involves a constant $c_s$
\begin{eqnarray}
 D(x) e^{- U(x) }  \frac{d O(x)}{d x} = c_s
\label{adjointzerofirst}
\end{eqnarray}
while the second integration leads to the linear combination 
\begin{eqnarray}
  O(x) = c_{0} l_0(x) + c_s s(x)
\label{CLOsat}
\end{eqnarray}
of two independent solutions where 
\begin{eqnarray}
l_0(x) && = 1 \ \ \text{  is associated to the conservation of the total probability (Eq. \ref{conservTotal})} 
\label{l0unity}
\end{eqnarray}
while 
\begin{eqnarray}
s(x)  =  \int_0^x dy \frac{ e^{U(y) }  } { D(y) } \ \ \ \text{ with its positive derivative} \ \ s'(x)=\frac{ e^{U(x) }  } { D(x) }
\label{Sx}
\end{eqnarray}
is called the 'scale function'
in the mathematical literature \cite{BookKarlin,BookYor,HandBook,BookBorodin}.
For instance for the Brownian motion corresponding to the vanishing force $F(x)=0$ and to the diffusion coefficient
$D(x)=1$, the potential vanishes $U(x)=0$ and the scale function of Eq. \ref{Sx} reduces to 
$s(x)  = x $, i.e. the averaged value of the position is the second conserved observable beyond the probability normalization associated to $l_0(x)=1$.

So in the general case, the final answer for the existence of conserved observables $O(x)$ 
via the dynamics of Eq. \ref{Oavdynfull}
depends on the boundary conditions.
In particular $l_0(x)=1$ will be conserved if the boundary conditions conserve the total probability as discussed around Eq. \ref{conservTotal}.


\subsection{ Link with the factorization of the adjoint $ {\cal F}^{\dagger} =\frac{d}{d m(x) } \frac{d}{d s(x) } $ in terms of the speed-measure $m(x)$
and the scale function $s(x)$ of the mathematical literature}

\label{subsec_scalespeed}

In the mathematical literature \cite{BookKarlin,BookYor,HandBook,BookBorodin}, the equilibrium solution of Eq. \ref{rhoeq}
is called the 'speed measure density' $m'(x)$ 
\begin{eqnarray}
\rho_{eq}(x)  = e^{- U(x)} = m'(x) 
\label{rhoeqmx}
\end{eqnarray}
while the non-equilibrium solution of Eq. \ref{rhoneq} can be then rewritten as the product between
 $m'(x)$ and the scale function $s(x)$ of Eq. \ref{Sx}
\begin{eqnarray}
\rho_{neq}(x) && = e^{- U(x)} \int_0^x dy \frac{ e^{U(y) }  } { D(y) } = m'(x) s(x)
\label{rhoneqms}
\end{eqnarray}

The adjoint operator of Eq. \ref{FPgeneratorprodadjoint} with Eq. \ref{defJopUadjoint} reads in terms of $m'(x)$ of Eq. \ref{rhoeqmx}
and $s'(x)$ of Eq. \ref{Sx}
\begin{eqnarray}
  {\cal F}^{\dagger} &&  =  {\cal J}^{\dagger}  \frac{\partial}{\partial x}
   = e^{ U(x) }   \frac{\partial}{\partial x} D(x) e^{- U(x) } \frac{\partial}{\partial x}
   \nonumber \\
   && = \left( \frac{ 1}{m'(x) }  \frac{\partial}{\partial x} \right) \left( \frac{ 1}{s'(x) }  \frac{\partial}{\partial x} \right)
  \label{FPgeneratorprodadjointUmath}
\end{eqnarray}
and is usually written in the mathematical literature \cite{BookKarlin,BookYor,HandBook,BookBorodin} as
\begin{eqnarray}
  {\cal F}^{\dagger}   =  \frac{d}{d m(x) } \frac{d}{d s(x) } 
  \label{FPgeneratorprodadjointmath}
\end{eqnarray}
while
the Fokker-Planck generator $ {\cal F} $ corresponds to
\begin{eqnarray}
 {\cal F} && =\frac{\partial}{\partial x}  D(x) e^{- U(x) }  \frac{\partial}{\partial x}  e^{ U(x) }  
\nonumber \\
&&  = \frac{\partial}{\partial x}   \frac{ 1}{s'(x) } \frac{\partial}{\partial x}  \frac{ 1}{m'(x) } 
\label{fokkerplanckgeneratorfactorizedUmath}
\end{eqnarray}

To be more concrete, let us now describe two physical interpretations of the scale function $s(x)$
and of the speed measure $m(x)$.


\subsubsection{ Interpretation of the scale function $s(x)$ via the exit probabilities $E^{L,R}(x)$ as a function of the initial condition $x$} 

When the diffusion process starts at $x$, the probability $E^{R}(x)$ to reach the right boundary $x_R$
before the left boundary $x_L$, 
and the complementary
probability $E^{L}(x)= 1- E^{R}(x)$ to reach the left boundary $x_L$ before the right boundary $x_R$,
 can be written in terms of the scale function $s(x)$
\begin{eqnarray}
E^{R}(x) = \frac{ s(x)-s(x_L) }{ s(x_R)-s(x_L) } = 1- E^{L}(x)
  \label{ExitRx}
\end{eqnarray}
Indeed, as explained in textbooks \cite{gardiner,vankampen,risken},
the probability $E^{R}(x) $ satisfies Eq. \ref{adjointzero} concerning conserved observables $O(x)$
 \begin{eqnarray}
0= {\cal F}^{\dagger} E^{R}(x) 
\label{adjointzeroExit}
\end{eqnarray}
and can be thus written as a linear combination of Eq. \ref{CLOsat}
involving the two independent solutions $l_0(x)=1$ and $s(x)$,
where the two constants are determined by the two boundary conditions at $X=x_R$ and $x=x_L$
 \begin{eqnarray}
 E^{R}(x_R) && =1
 \nonumber \\
 E^{R}(x_L) && =0
\label{BCExit}
\end{eqnarray}

The exit probability $E^{R}(x) $ can be decomposed
\begin{eqnarray}
E^{R}(x) = \int_0^{+\infty} dt \pi^R_t(x)
  \label{ExitRxt}
\end{eqnarray}
in terms of the probability $\pi^R_t(x) $ to reach the right boundary $x_R$ at the time $t$ without having visited the other boundary $x_L$,
whose dynamics is governed by the adjoint operator ${\cal F}^{\dagger} $ \cite{gardiner,vankampen,risken}
 \begin{eqnarray}
\partial_t \pi^R_t(x)= {\cal F}^{\dagger} \pi^R_t(x)
\label{adjointzeroExitt}
\end{eqnarray}
while the two boundary conditions at $X=x_R$ and $x=x_L$ read
 \begin{eqnarray}
 \pi^R_t(x_R) && =\delta(t)
 \nonumber \\
 \pi^R_t(x_L) && =0
\label{BCExitt}
\end{eqnarray}

Equivalently, one can consider the cumulative distribution
\begin{eqnarray}
E^{R}_t(x) \equiv \int_0^t d\tau \pi^R_{\tau}(x)  
  \label{ExitRxtt}
\end{eqnarray}
whose dynamics is also governed by the adjoint operator ${\cal F}^{\dagger} $
\begin{eqnarray}
{\cal F}^{\dagger}  E^{R}_t(x) && =  \int_0^t d\tau {\cal F}^{\dagger}\pi^R_{\tau}(x)
=  \int_0^t d\tau \partial_{\tau} \pi^R_{\tau}(x) =  \pi^R_t(x)
\nonumber \\
\partial_t E^{R}_t(x) && =   \pi^R_t(x) = {\cal F}^{\dagger}  E^{R}_t(x)
  \label{ExitRxttdyn}
\end{eqnarray}
with the boundary conditions obtained from Eq. \ref{BCExitt}
 \begin{eqnarray}
E^{R}_t(x_R) && = 1
 \nonumber \\
 E^R_t(x_L) && =0
\label{BCExittCumul}
\end{eqnarray}
and the initial condition at $t=0$
\begin{eqnarray}
E^{R}_{t =0} (x) &&  = 0  \text { for $x_L \leq x <x_R$ }
\nonumber \\
E^{R}_{t =0} (x_R) && =1
  \label{ExitRxt0}
\end{eqnarray}
while the limit for $t \to + \infty$ corresponds to
the exit probability $E^{R}(x) $ of Eq. \ref{ExitRx} making use of Eq. \ref{ExitRxt}
\begin{eqnarray}
E^{R}_{t =+\infty} (x) =\int_0^{+\infty} dt \pi^R_t(x) = E^{R}(x) = \frac{ s(x)-s(x_L) }{ s(x_R)-s(x_L) } 
  \label{ExitRxtinfty}
\end{eqnarray}


\subsubsection{ Interpretation of the speed measure $m(x)$ via the cumulative equilibrium distribution $C_{eq}(x)$} 

The derivative $m'(x)=e^{- U(x)}$ of Eq. \ref{rhoeqmx}
yields that the normalized equilibrium steady state reads
\begin{eqnarray}
P_{eq}(x)  = \frac{ e^{- U(x)} }{ \int_{x_L}^{x_R} dy e^{- U(y)}} = \frac{ m'(x) }{ m(x_R)-m(x_L) }
\label{Peqmx}
\end{eqnarray}
As a consequence, the corresponding cumulative equilibrium distribution $C_{eq}(x)$ can be written in terms of the speed measure $m(.)$ alone
\begin{eqnarray}
C_{eq}(x) \equiv \int_{x_L}^x dz P_{eq}(z)  = \frac{ m(x) -m(x_L)}{ m(x_R)-m(x_L) }
\label{Ceqmx}
\end{eqnarray}

For later purposes,
 it is useful to introduce 
 the time-dependent cumulative distribution
\begin{eqnarray}
C_t(x \vert x_R) \equiv \int_{x_L}^x dz P_t(z \vert x_R)  
\label{Cumultx}
\end{eqnarray}
satisfying the boundary conditions
\begin{eqnarray}
C_t(x_R \vert x_R) && = \int_{x_L}^{X_R} dz P_t(z \vert x_R)  =1
\nonumber \\
C_t(x_L \vert x_R) && =0
\label{CumultxBC}
\end{eqnarray}
and the initial condition
\begin{eqnarray}
C_{t=0}(x \vert x_R) &&  = 0  \text { for $x_L \leq x <x_R$ }
\nonumber \\
C_{t=0}(x_R \vert x_R) && =1
  \label{Ct0}
\end{eqnarray}
while its dynamics 
\begin{eqnarray}
\partial_t C_t(x\vert x_R) 
\equiv \int_{x_L}^x dz \partial_t P_t(z\vert x_R)  = - \int_{x_L}^x dz \frac{ \partial }{\partial z} J_t(z\vert x_R) 
= J_t(x_L\vert x_R) - J_t(x\vert x_R) = - J_t(x\vert x_R)
\label{Cumultxderi}
\end{eqnarray}
is governed by the current $J_t(x\vert x_R) $ as a consequence of the vanishing-current B.C. $J_t(x_L) =0 $ of Eq. \ref{vanishinhJ}, so that the integration over the time-window $[0,t]$ 
yields using the vanishing initial condition for $x_L \leq x <x_R $
\begin{eqnarray}
C_t(x\vert x_R)  = -  \int_0^t d \tau J_{\tau} (x\vert x_R) \ \ \ \text { for $x_L \leq x <x_R$ }
\label{JtCumultxderiinteg}
\end{eqnarray}
Let us anticipate a little bit, and use that the dynamics of the current $J_t(x)$ is governed by the opeartor 
${\hat {\cal F} }   \equiv  - {\cal J}  \frac{\partial}{\partial x}$ as will be discussed in Eq. \ref{dyncurrent}
of the next section, in order to obtain that the dynamics of $C_t(x\vert x_R)  $ is governed by the same operator
using Eqs \ref{Cumultxderi}
and \ref{JtCumultxderiinteg}
\begin{eqnarray}
{\hat {\cal F} } C_t(x\vert x_R)  && =-  \int_0^t d \tau {\hat {\cal F} } J_{\tau} (x\vert x_R) 
=-  \int_0^t d \tau \partial_{\tau} J_{\tau} (x\vert x_R) = J_t(x\vert x_R)
\nonumber \\
\partial_t C_t(x\vert x_R) 
&& = - J_t(x\vert x_R) = {\hat {\cal F} } C_t(x\vert x_R) 
\label{JtCumultxderiintegdyn}
\end{eqnarray}
while the limit for $t \to + \infty$ corresponds to
the 
cumulative equilibrium distribution $C_{eq}(x)$ of Eq. \ref{Ceqmx}
\begin{eqnarray}
C_{t =+ \infty} (x\vert x_R) = C_{eq}(x)  = \frac{ m(x) -m(x_L)}{ m(x_R)-m(x_L) }
\label{Ceqmxtinfty}
\end{eqnarray}


\subsection{ Similarity transformation between the Fokker-Planck generator $ {\cal F}$ and the
hermitian supersymmetric quantum Hamiltonian ${\cal H}  \equiv - e^{ \frac{U(x)}{2} }  {\cal F}  e^{- \frac{U(x)}{2} }= Q^{\dagger}  Q $}

The fully factorized form of Eq. \ref{fokkerplanckgeneratorfactorizedU}
for the Fokker-Planck generator $  {\cal F}$
directly leads to the standard similarity transformation towards
the hermitian quantum Hamiltonian
\begin{eqnarray}
{\cal H} && \equiv - e^{ \frac{U(x)}{2} }  {\cal F}  e^{- \frac{U(x)}{2} } 
= -  e^{ \frac{U(x)}{2} }  \frac{\partial}{\partial x} e^{- \frac{U(x)}{2} } D(x) e^{ - \frac{U(x)}{2} }\frac{\partial}{\partial x}  e^{ \frac{U(x)}{2} }
 \equiv Q^{\dagger} Q
\label{Hsusy}
\end{eqnarray}
with its well-known supersymmetric factorization in terms of the first order operators $Q$ and $Q^{\dagger}$ 
\begin{eqnarray}
Q && \equiv     - \sqrt{D(x)} e^{ - \frac{U(x)}{2} }  \frac{\partial}{\partial x} e^{ \frac{U(x)}{2} }
\nonumber \\
 Q^{\dagger} && 
 = e^{ \frac{U(x)}{2} } \frac{\partial}{\partial x } e^{- \frac{U(x)}{2} } \sqrt{D(x)}
\label{HsusyQ}
\end{eqnarray}
This supersymmetric quantum Hamiltonian (see the reviews \cite{review_susyquantum,Mielnik_review,sasaki_reviewSusyQM,Review_factorization}
with various scopes on supersymmetric quantum mechanics)
has been used extensively to analyze the diffusion processes with various boundary conditions (see for instance \cite{jpb_antoine,pierre,texier,c_lyapunov,us_gyrator,us_kemeny,c_pearson,c_boundarydriven}).


\subsection{ Spectral decomposition of the Fokker-Planck propagator }

\label{subsec_spectralP}

For the vanishing-current B.C. of Eq. \ref{vanishinhJ} leading to the 
vanishing steady current $J_{eq}=0$ (detailed-balance) and to the equilibrium steady state of Eq. \ref{rhoeq},
the similarity transformation described in the previous subsection has
important consequences for the spectral analysis of the relaxation towards equilibrium as described in textbooks \cite{gardiner,vankampen,risken}.
Let us assume that the real eigenvalues of the quantum Hamiltonian ${\cal H}$
are only discrete $E_n$ labelled by $n=0,1,..,+\infty$, so that 
the spectral decomposition of the Euclidean quantum propagator reads
\begin{eqnarray}
\psi_t(x \vert x_0)  \equiv  \langle x \vert e^{- {\cal H} t} \vert x_0 \rangle
= \sum_{n=0}^{+\infty} e^{- t E_n}  \phi_n (x)  \phi_n (x_0) 
\label{psispectral}
\end{eqnarray}
where the corresponding eigenstates $\phi_n(x)$ of the hermitian quantum Hamiltonian ${\cal H}$
\begin{eqnarray}
E_n \phi_n(x) = {\cal H}   \phi_n(x) 
\label{hsusyeigen}
\end{eqnarray}
have been chosen to be real to simplify the notations, and 
satisfy the orthonormalization
\begin{eqnarray}
\delta_{n n' } = \langle \phi_n \vert \phi_{n'} \rangle 
= \int_{x_L}^{x_R}  dx \phi_n(x) \phi_{n'}(x)
\label{orthophin}
\end{eqnarray}
The positive quantum ground-state $\phi_0(x)$ associated to the vanishing energy $E_0=0$
is then annihilated by the first-order operator $Q$ and is simply the square-root of the equilibrium steady state $P_{eq}(x)$
\begin{eqnarray}
  \phi_0(x) =   \sqrt{ P_{eq}(x) } = \frac{ e^{ - \frac{ U(x)}{2} } }{\sqrt { \int_{x_L}^{x_R} dy e^{-U(y)}} }
 \label{phi0}
\end{eqnarray}

The corresponding spectral decomposition 
of the Fokker-Planck propagator $P_t(x \vert x_0) $
reads
via the similarity transformation of Eq. \ref{Hsusy}
\begin{eqnarray}
P_t(x \vert x_0)  \equiv \langle x \vert e^{ {\cal F} t} \vert x_0 \rangle = \sqrt{ \frac{P_{eq}(x)}{P_{eq}(x_0) }  } \psi_t(x \vert x_0) 
&& = \sum_{n=0}^{+\infty} e^{- t E_n} \bigg(\sqrt{ P_{eq}(x) } \langle x \vert \phi_n \rangle \bigg) 
\bigg( \langle\phi_n \vert x_0 \rangle \frac{1}{ \sqrt{ P_{eq}(x_0) }}\bigg) 
\nonumber \\
&& = \sum_{n=0}^{+\infty} e^{- t E_n}  r_n(x) l_n(x_0)
= P_{eq}(x) + \sum_{n=1}^{+\infty} e^{- t E_n}  r_n(x) l_n(x_0) 
\label{FPspectralq}
\end{eqnarray}
where
\begin{eqnarray}
 l_n ( x )  && \equiv \phi_n(x) \frac{1}{ \sqrt{ P_{eq}(x) }}
\nonumber \\
r_n(x) && \equiv \sqrt{ P_{eq}(x) } \phi_n(x) = P_{eq}(x) l_n(x) 
\label{rnlnphin}
\end{eqnarray}
are the left eigenvectors $l_n(x)$
and the right eigenvectors $r_n(x)$ of the Fokker-Planck generator ${\cal F} $ associated to the eigenvalue $(-E_n)$
\begin{eqnarray}
  -E_n l_n(x) && = {\cal F}^{\dagger} l_n(x)   
 \nonumber \\
 -E_n r_n(x) &&= {\cal F} r_n(x)  
 \label{fokkerplanckeigen}
\end{eqnarray}
that satisfy the bi-orthonormalization relations
\begin{eqnarray}
\delta_{n n' } = \langle l_n \vert r_{n'} \rangle = \int_{x_L}^{x_R} dx l_n(x) r_{n'}(x)
 \label{ortholr}
\end{eqnarray}
The vanishing eigenvalue $E_0=0$ is associated to the convergence towards the steady state 
$P_{eq}(x) $ for any initial condition $x_0$
\begin{eqnarray}
  r_0(x) &&= P_{eq}(x)
 \nonumber \\
  l_0(x) && =    1
 \label{r0l0}
\end{eqnarray}

The vanishing-current boundary conditions $  J_t(x_L)=0=J_t(x_R)$ for the current
${\cal J} P_t(x \vert x_0) $ associated to the probability $ P_t(x \vert x_0)$ with a given initial condition
\begin{eqnarray}
 {\cal J} P_t(x \vert x_0)   
 =  \sum_{n=0}^{+\infty} e^{- t E_n} \bigg( { \cal J } r_n(x) \bigg) l_n(x_0)
\label{SpectralJt}
\end{eqnarray}
mean that the currents $j_n(x)$ associated to the right eigenvector $r_n(x)$
\begin{eqnarray}
  j_n(x) && \equiv {\cal J} r_n(x) = \bigg( F(x)  - D (x)  \partial_{x} \bigg) r_n(x) 
  = - D(x) \bigg(  U'(x) r_n(x) +  r_n'(x) \bigg) 
 \label{jnright}
\end{eqnarray}
that can also be rewritten either in terms of the quantum eigenstate $\phi_n(x)$
\begin{eqnarray}
  j_n(x)  &&  
 = - \sqrt{ P_{eq}(x)} D(x)    \bigg(  \frac{U'(x)}{2}  \phi_n(x)+ \phi_n'(x) \bigg)    
 \label{jnquantum}
\end{eqnarray}
or in terms of the left eigenvectors $l_n(x)$ of Eq. \ref{rnlnphin}
\begin{eqnarray}
  j_n(x)    = - P_{eq}(x) D(x) l_n'(x) 
 \label{jnleft}
\end{eqnarray}
should all vanish at the two boundaries
\begin{eqnarray}
  j_n(x_L) = j_n(x_R) =0 
  = P_{eq}(x_L) D(x_L) l_n'(x_L)  = P_{eq}(x_R) D(x_R) l_n'(x_R) 
 \label{jnrightzero}
\end{eqnarray}

As a consequence, to describe the convergence towards the equilibrium state $P_{eq}(x)$,
it is often simpler to rewrite the spectral decomposition of Eq. \ref{FPspectralq}
\begin{eqnarray}
P_t(x \vert x_0)  = P_{eq}(x) \left[ \sum_{n=0}^{+\infty} e^{- t E_n}  l_n(x) l_n(x_0) \right]
= P_{eq}(x) \left[ 1 + \sum_{n=1}^{+\infty} e^{- t E_n}  l_n(x) l_n(x_0) \right]
\label{FPspectralqleft}
\end{eqnarray}
in terms of the left eigenvectors $l_n(x)$ alone, since they satisfy the simpler boundary conditions of Eq. \ref{jnrightzero},
while the bi-orthonormalization of Eq. \ref{ortholr} translates into the orthogonal family property for the left eigenvectors $l_n(x)$ with respect to the equilibrium measure $P_{eq}(x)$
\begin{eqnarray}
\delta_{n n' } =  \int_{x_L}^{x_R} dx l_n(x) r_{n'}(x) =  \int_{x_L}^{x_R} dx l_n(x) l_{n'}(x) P_{eq}(x)
 \label{ortholrl}
\end{eqnarray}

\subsection{ Discussion } 

In this section, we have revisited the well-known properties of reversible Fokker-Planck generators governing the 
dynamics for the probability $P_t(x)$ alone, via the unifying and clarifying perspective of their factorized structure into two first-order differential generators.
In the next section, the goal is to analyze similarly the properties of the dynamics of  
the current $J_t(x)$ alone.


\section{ Supersymmetric partner ${\hat {\cal F} }= - {\cal J}  \frac{\partial}{\partial x}$ governing the dynamics of the current $J_t(x)$  }

\label{sec_diffusionJ}

In this section, we describe how the dynamics of the
current $J_t(x)$ is governed by the supersymmetric partner 
${\hat {\cal F} }   \equiv  - {\cal J}  \frac{\partial}{\partial x}$ of 
the Fokker-Planck generator ${\cal F}  =  -  \frac{\partial}{\partial x} {\cal J}  $
discussed in the previous section, and we analyze the consequences
for the relations between their spectral properties.

\subsection{ Dynamics of the current $J_t(x)$ governed by the supersymmetric partner 
${\hat {\cal F} }   \equiv  - {\cal J}  \frac{\partial}{\partial x}$ of ${\cal F}  =  -  \frac{\partial}{\partial x} {\cal J}  $ }

The dynamics of the current $J_t(x)$ obtained from Eqs 
\ref{continuity} and \ref{fokkerplanckcurrent}
\begin{eqnarray}
 \partial_t J_t( x ) = {\cal J } \partial_t P_t( x )    
    =  - {\cal J}  \frac{\partial}{\partial x}  J_t(x) \equiv {\hat {\cal F} } J_t(x)
\label{dyncurrent}
\end{eqnarray}
is governed by the 
supersymmetric partner 
of the Fokker-Planck generator ${\cal F}  =  -  \frac{\partial}{\partial x} {\cal J}  $,
i.e. the two first-order differential operators appear in the opposite order
\begin{eqnarray}
{\hat {\cal F} } &&  \equiv  - {\cal J}  \frac{\partial}{\partial x} 
= - \left(  F(x)   - D(x) \frac{\partial}{\partial x}  \right)\frac{\partial}{\partial x}
\nonumber \\&& 
= -   F(x)   \frac{\partial}{\partial x}+  D(x) \frac{\partial^2}{\partial x^2} 
\label{susypartnerFP}
\end{eqnarray}
with the alternative fully factorized expression using Eq. \ref{defJopU}
involving the potential $U(x)$
\begin{eqnarray}
{\hat {\cal F} } &&  \equiv  - {\cal J}  \frac{\partial}{\partial x} 
= D(x) e^{- U(x) }  \frac{\partial}{\partial x}  e^{ U(x) }  \frac{\partial}{\partial x} 
\label{susypartnerFPU}
\end{eqnarray}

With the mathematical notations of Eqs \ref{FPgeneratorprodadjointUmath}
and \ref{FPgeneratorprodadjointmath},
the supersymmetric partner reads
\begin{eqnarray}
{\hat {\cal F} } && 
= \left( \frac{ 1}{s'(x) }  \frac{\partial}{\partial x} \right)  \left( \frac{ 1}{m'(x) }  \frac{\partial}{\partial x} \right)
= \frac{d}{d s(x) } \frac{d}{d m(x) }
\label{susypartnerFPUmath}
\end{eqnarray}
that can be considered as the operator obtained by exchanging the roles of 
the speed measure $m(x)$ and the scale function $s(x)$
with respect to the adjoint operator $
  {\cal F}^{\dagger}   =  \frac{d}{d m(x) } \frac{d}{d s(x) } $
  of Eq. \ref{FPgeneratorprodadjointmath}, as also considered in Eqs (17-18) of the related recent mathematical work \cite{mathSusyScaleSpeed}.

  Note that the vanishing-current Boundary Conditions $J_t(x_L) =0 = J_t(x_R) $
of Eq. \ref{vanishinhJ} correspond to 'Absorbing Boundary Conditions' for the current $J_t(x)$
and lead to the vanishing steady value $J_{eq}=0$.


\subsection{  Intertwining relations between the supersymmetric partners ${\cal F}   =  -  \frac{\partial}{\partial x} {\cal J} $ and $ {\hat {\cal F} }  = - {\cal J}  \frac{\partial}{\partial x}$ with consequences for their eigenvalues and eigenvectors}

\subsubsection{  Intertwining relations involving the current operator $ {\cal J} $ with consequences for their right eigenvectors}

The two supersymmetric partners ${\cal F}   =  -  \frac{\partial}{\partial x} {\cal J} $ and $ {\hat {\cal F} }  = - {\cal J}  \frac{\partial}{\partial x}$
satisfy the intertwining relation involving the current operator ${\cal J} $ 
\begin{eqnarray}
{\cal J}{\cal F}   &&=  - {\cal J} \frac{\partial}{\partial x} {\cal J} =  {\hat {\cal F} } {\cal J}
  \label{interwJ}
\end{eqnarray}
The eigenvalue equation of Eq. \ref{fokkerplanckeigen} 
for the right eigenvectors $l_n(x)$ 
of the Fokker-Planck generator ${\cal F} $ 
\begin{eqnarray}
-E_n r_n(x) =  {\cal F} r_n(x) =  -  \frac{\partial}{\partial x} {\cal J} r_n(x)
  \label{fokkerplanckeigenr}
\end{eqnarray}
can be thus split into two first-order equations
\begin{eqnarray}
  j_n(x) && \equiv {\cal J} r_n(x)
  \nonumber \\
   \frac{\partial}{\partial x} j_n(x)  && =   E_n r_n(x) 
  \label{splittingr}
\end{eqnarray}
where $j_n(x)={\cal J} r_n(x)$ is a right eigenvector of the supersymmetric partner $ {\hat {\cal F} } $
associated to the eigenvalue $(-E_n)$
\begin{eqnarray}
{\hat {\cal F} } j_n(x)  && =  - {\cal J}  \frac{\partial}{\partial x} {\cal J} r_n(x) = {\cal J}{\cal F} r_n(x) = - E_n{\cal J} r_n(x)
= -E_n j_n(x)
  \label{eigenjn}
\end{eqnarray}
except for $n=0$ where the steady current $j_0={\cal J} r_0={\cal J}P_{eq}=0$ associated to the equilibrium steady state $P_{eq}(x)$ vanishes.

Note that the vanishing-current boundary conditions $0=  j_n(x_L) = j_n(x_R) $ of Eq. \ref{jnrightzero}
involve the first part $ j_n(x)  \equiv {\cal J} r_n(x) $ of the splitting of Eq. \ref{splittingr}
that give the correspondence between the B.C. for the $j_n(x)$ and for the $r_n(x)$
\begin{eqnarray}
  j_n(x)  \equiv {\cal J} r_n(x)  =0 \text{ for $x=x_L$ and for $x=x_R$}
  \label{jnzeroagain}
\end{eqnarray}


\subsubsection{  Intertwining relations involving the derivative operator $\frac{\partial}{\partial x} $ with consequences for their left eigenvectors}

The two supersymmetric partners ${\cal F}   =  -  \frac{\partial}{\partial x} {\cal J} $ and $ {\hat {\cal F} }  = - {\cal J}  \frac{\partial}{\partial x}$
satisfy the intertwining relation 
involving the derivative operator $\frac{\partial}{\partial x} $
\begin{eqnarray}
{\cal F} \frac{\partial}{\partial x}  
&&=  -  \frac{\partial}{\partial x} {\cal J} \frac{\partial}{\partial x} = \frac{\partial}{\partial x}{\hat {\cal F} } 
  \label{interwderivative}
\end{eqnarray}
or equivalently for their adjoint operators
\begin{eqnarray}
 \frac{\partial}{\partial x}   {\cal F}^{\dagger} 
&&=    \frac{\partial}{\partial x} {\cal J}^{\dagger} \frac{\partial}{\partial x} = {\hat {\cal F} }^{\dagger}  \frac{\partial}{\partial x}
  \label{interwderivativeadjoint}
\end{eqnarray}

The eigenvalue equation of Eq. \ref{fokkerplanckeigen} 
for the left eigenvectors $l_n(x)$ 
of the Fokker-Planck generator ${\cal F} $ 
\begin{eqnarray}
  -E_n l_n(x) && = {\cal F}^{\dagger} l_n(x)   =  {\cal J}^{\dagger } \frac{\partial}{\partial x} l_n(x)
 \label{fokkerplanckeigenln}
\end{eqnarray}
can be thus split into two first-order equations
\begin{eqnarray}
 l_n  (x) && = {\cal J}^{\dagger}    i_n ( x)
\nonumber \\
 E_n    i_n ( x)  && =- \frac{\partial}{\partial x}  l_n ( x) 
\label{splittingrl}
\end{eqnarray}
where $i_n(x)=- \frac{l_n'(x)}{E_n} $ is a left eigenvector of the supersymmetric partner $ {\hat {\cal F} } $
associated to the eigenvalue $(-E_n)$
\begin{eqnarray}
{\hat {\cal F} }^{\dagger} i_n(x)  && = \frac{\partial}{\partial x}  {\cal J}^{\dagger }  i_n(x)
= \frac{\partial}{\partial x}  l_n  (x) = -E_n i_n(x)
 \label{eigenin}
\end{eqnarray}
except for $n=0$ where $l_0'(x)=0$ and $E_0=0$ vanishes.

Note that the boundary conditions of Eq. \ref{jnrightzero} for the left eigenvectors $l_n(x)$
translate for the eigenvectors $i_n(x)$ of Eq. \ref{splittingrl}
into
\begin{eqnarray}
  P_{eq}(x) D(x) i_n(x) = 0  \text{ for $x=x_L$ and for $x=x_R$}
 \label{Oavdynfulllnbci}
\end{eqnarray}


\subsubsection{Conclusion for the spectral decomposition of the propagator $\langle x \vert e^{ t {\hat {\cal F} }} \vert x_0 \rangle $ 
associated to the supersymmetric partner $ {\hat {\cal F} }$ }

Using the vanishing-current boundary conditions and the eigenvalue Eq. \ref{fokkerplanckeigenr},
one obtains that the right eigenvectors $j_m(x)$ of Eq. \ref{eigenjn}
and the left eigenvectors $i_n(x)$ of Eq. \ref{eigenin}
satisfy the orthonormalization inherited from Eq. \ref{ortholr}
\begin{eqnarray}
\langle i_n \vert j_m \rangle && = \int_{x_L}^{x_R} i_n(x) j_m(x) 
=  \int_{x_L}^{x_R} \bigg( - \frac{1}{ E_n } \frac{\partial}{\partial x}  l_n ( x) \bigg)  j_m(x)
= \frac{1}{ E_n }  \int_{x_L}^{x_R}   l_n ( x)  \frac{\partial}{\partial x} j_m(x)
= \frac{1}{ E_n }  \int_{x_L}^{x_R}   l_n ( x)  \frac{\partial}{\partial x} {\cal J} r_m(x)
\nonumber \\
&& =  \frac{1}{ E_n }  \int_{x_L}^{x_R}   l_n ( x)  (- {\cal F} ) r_m(x)
= \frac{E_m }{ E_n } \int_{x_L}^{x_R}   l_n ( x)   r_m(x) =  \frac{E_m }{ E_n }\langle l_n \vert r_m \rangle = \delta_{n,m}
\label{orthoijn}
\end{eqnarray}
As a consequence, the spectral decomposition of the propagator $\langle x \vert e^{ t {\hat {\cal F} }} \vert x_0 \rangle $ 
associated to the supersymmetric partner $ {\hat {\cal F} }$ describing the convergence towards zero-current
\begin{eqnarray}
\langle x \vert e^{ t {\hat {\cal F} }} \vert x_0 \rangle 
&&=  \sum_{n=1}^{+\infty} e^{- t E_n}  j_n(x) i_n(x_0)
  \label{spectralPartner}
\end{eqnarray}
involves the same non-vanishing eigenvalues $E_n>0$ as Eq. \ref{FPspectralq},
while their corresponding right and left eigenvectors are related via Eqs \ref{splittingr}
and \ref{splittingrl}
respectively.


\subsection{ Link with the supersymmetric partner ${\breve H} = Q Q^{\dagger} $ 
of the quantum Hamiltonian $H=Q^{\dagger} Q $} 

Since we have written in Eq. \ref{Hsusy} the similarity transformation between the Fokker-Planck generator ${\cal F}$
and the quantum Hamiltonian $H=Q^{\dagger} Q $, 
it is natural to see how the supersymmetric partner
${\hat F}$ defined in Eq. \ref{susypartnerFPU}
is related to the supersymmetric partner ${\breve H} = Q Q^{\dagger} $ of the quantum Hamiltonian
\begin{eqnarray}
{\breve H} =Q Q^{\dagger} =     - \sqrt{D(x)} e^{ - \frac{U(x)}{2} }  \frac{\partial}{\partial x} e^{ U(x) } \frac{\partial}{\partial x } e^{- \frac{U(x)}{2} } \sqrt{D(x)}
\label{Hsusybreve}
\end{eqnarray}

One obtains that they are related via the similarity transformation
\begin{eqnarray}
{\breve H}  = - \frac{1}{e^{- \frac{U(x)}{2} } \sqrt{D(x)} }{\hat {\cal F} } e^{- \frac{U(x)}{2} } \sqrt{D(x)} 
\label{susypartnerFPUbiss}
\end{eqnarray}
and that the ground state of ${\breve H} $ 
\begin{eqnarray}
{\breve \psi}_{0} (x) \propto \frac{e^{\frac{U(x)}{2} }}{\sqrt{D(x)}}   \equiv e^{ - \frac{ {\breve U}(x) }{2} }
\label{susypartnerbreveHGS}
\end{eqnarray}
involves the potential 
\begin{eqnarray}
{\breve U}(x) = - U(x) +  \ln D(x) 
 \label{Uxbreve}
\end{eqnarray}
that will also appear as $ {\mathring U}(x)$ in Eq. \ref{Uxring}
via the construction described in the next section.


\subsection{ Discussion }

In this section, we have analyzed the properties of the dynamics of  
the current $J_t(x)$ alone.
In contrast to the normalized positive probability $P_t(x)$ that evolves with the Fokker-Planck generator ${\cal F}$
and converges towards the equilibrium state $P_{eq}(x)$, 
 the current $J_t(x)$ evolving with the supersymmetric partner ${\hat {\cal F} } $ 
 can be either positive or negative, is not normalized and converges towards zero.
 Nevertheless, it is interesting to try to re-interpret the supersymmetric partner 
 ${\hat {\cal F} } = - {\cal J}  \frac{\partial}{\partial x}$
in terms of the initial Fokker-Planck generator with different parameters
as discussed in the two next sections concerning two different re-interpretations.


\section{ Interpreting the supersymmetric partner ${\hat {\cal F} } = - {\cal J}  \frac{\partial}{\partial x}$ 
as the adjoint ${\mathring {\cal F}}^{\dagger} ={\mathring {\cal J} }^{\dagger} \frac{\partial}{\partial x}  $ of  the Fokker-Planck generator 
${\mathring {\cal F}}=-  \frac{\partial}{\partial x}  {\mathring {\cal J} }$
associated to some dual force ${\mathring F}(x) $}

\label{sec_duality}

In this section, we describe how the supersymmetric partner ${\hat {\cal F} } = - {\cal J}  \frac{\partial}{\partial x}$
described in the previous section can be interpreted as the adjoint ${\mathring {\cal F}}^{\dagger} ={\mathring {\cal J} }^{\dagger} \frac{\partial}{\partial x}  $ of  the Fokker-Planck generator 
${\mathring {\cal F}}=-  \frac{\partial}{\partial x}  {\mathring {\cal J} }$
associated to some dual force ${\mathring F}(x) $,
while the diffusion coefficient $D(x)$ remains the same.

\subsection{ Correspondence between ${\hat {\cal F} } \equiv - {\cal J}  \frac{\partial}{\partial x} $
and ${\mathring {\cal F}}^{\dagger}  \equiv {\mathring {\cal J} }^{\dagger} \frac{\partial}{\partial x}     $}

The supersymmetric partner ${\hat {\cal F} }  = - {\cal J}  \frac{\partial}{\partial x}$ of Eq. \ref{susypartnerFP}
contains the derivative operator $\frac{\partial}{\partial x} $ on the right
and can be thus identified 
 with
the adjoint ${\mathring {\cal F}}^{\dagger}  $ of  Eq. \ref{FPgeneratorprodadjoint}
involving the same diffusion coefficient $D(x)$ but another force ${\mathring F}(x) $
\begin{eqnarray}
  {\mathring {\cal F}}^{\dagger}  \equiv {\mathring {\cal J} }^{\dagger} \frac{\partial}{\partial x} 
  \ \ \text{ with } \ \ 
  {\mathring {\cal J} }^{\dagger} \equiv \left[{\mathring F}(x)+D'(x)\right]  + D(x) \frac{\partial}{\partial x}  
  \label{FPgeneratorprodadjointring}
\end{eqnarray}
i.e. the two first-order current-operators ${\mathring {\cal J} }^{\dagger} $ and ${\cal J} $ should satisfy 
\begin{eqnarray}
 {\mathring {\cal J} }^{\dagger} =  - {\cal J} \equiv -   F(x)  +  D(x) \frac{\partial}{\partial x} 
\label{dualityringcurrent}
\end{eqnarray}
so that the dual force ${\mathring F}(x) $ reads
\begin{eqnarray}
 {\mathring F}(x)   = - F(x) -  D'(x)
\label{dualityringforce}
\end{eqnarray}
The corresponding dual potential ${\mathring U} (x)$  associated to the dual force ${\mathring F}(x) $ via Eq. \ref{Ux}
has for derivative 
\begin{eqnarray}
{\mathring U}'(x)  &&= -  \frac{{\mathring F}(x)}{D(x)} 
=   \frac{F(x)}{D(x)}
 +  \frac{D'(x)}{D(x)}
\nonumber \\
 && = - U'(x) +  \frac{d}{dx} \ln D(x) 
 \label{Uxdiff}
\end{eqnarray}
and can be thus chosen to be via integration
\begin{eqnarray}
{\mathring U}(x) = - U(x) +  \ln D(x) 
 \label{Uxring}
\end{eqnarray}
so that one recovers the potential 
${\breve U}(x) $ of Eq. \ref{Uxbreve}
of the last section.
This duality has been discussed in detail for the case of boundary-driven B.C. in \cite{c_boundarydriven}
and in the sections concerning non-interacting particles in \cite{tailleurMapping},
while it is known as Siegmund duality for other B.C. as recalled in the next subsection.


\subsection{Link with the Siegmund duality }

The duality potential ${\mathring U}(x) $ of Eq. \ref{Uxring} 
yields that the corresponding speed-measure-density $ {\mathring m}'(x) $ via Eq. \ref{rhoeqmx}
and the corresponding scale-function density $ {\mathring s}'(x) $ via Eq. \ref{Sx}
\begin{eqnarray}
  {\mathring m}'(x)  \equiv e^{- {\mathring U}(x)} = \frac{ e^{U(x)}}{D(x)} =s'(x)
  \nonumber \\
  {\mathring s}'(x) \equiv \frac{ e^{{\mathring U}(x) }  } { D(x) } = e^{- U(x) } = m'(x)
\label{speedscalering}
\end{eqnarray}
are simply exchanged in the dual model with respect to the densities $[ m'(x), s'(x) ]$ associated to the potential $U(x)$
(see also the related recent mathematical work \cite{mathSusyScaleSpeed} with their Eqs (17-18)).

As a consequence,
the exit probability of Eq. \ref{ExitRx} for the dual model that involves its scale function ${\mathring s}(x)=m(x) $
\begin{eqnarray}
{\mathring E}^{R}(x) = \frac{ {\mathring s}(x)-{\mathring s}(x_L) }{ {\mathring s}(x_R)-{\mathring s}(x_L) } 
= \frac{ m(x) -m(x_L)}{ m(x_R)-m(x_L) } = C_{eq}(x) 
  \label{ExitRxring}
\end{eqnarray}
coincides with the cumulative distribution $C_{eq}(x)$ of Eq. \ref{Ceqmx}.
This property can be extented to an arbitrary time $t$ by considering their time-dependent counterparts 
that were discussed in detail in subsection \ref{subsec_scalespeed} :

(i) for the dual model, the dynamics of Eq. \ref{ExitRxttdyn}
for ${\mathring E}^{R}_t(x) $ of Eq. \ref{ExitRxtt}
involves the adjoint operator ${\mathring {\cal F}}^{\dagger} $
\begin{eqnarray}
\partial_t {\mathring E}^{R}_t(x) = {\mathring {\cal F}}^{\dagger}  {\mathring E}^{R}_t(x)
  \label{ExitRxttdynring}
\end{eqnarray}
with the boundary conditions of Eq. \ref{BCExittCumul}
and the initial condition of Eq. \ref{ExitRxt0}.

(ii) for the initial model, the cumulative distribution $C_t(x \vert x_R)$ of Eq. \ref{Cumultx}
satisfies the same boundary conditions of Eq. \ref{CumultxBC}
and the same initial condition of Eq. \ref{Ct0},
while its dynamics of Eq. \ref{JtCumultxderiintegdyn} is governed by the supersymmetric partner ${\hat {\cal F} } $
that coincides with ${\mathring {\cal F}}^{\dagger}  $ 
\begin{eqnarray}
\partial_t C_t(x\vert x_R)  = {\hat {\cal F} } C_t(x\vert x_R) = {\mathring {\cal F}}^{\dagger}  C_t(x\vert x_R)
\label{JtCumultxderiintegdyndualinter}
\end{eqnarray}

(iii) As a consequence, the two observables ${\mathring E}^{R}_t(x) $ and $ C_t(x\vert x_R)$
that satisfy the same dynamics with the same initial condition and the same boundary conditions
coincide
 \begin{eqnarray}
{\mathring E}^{R}_t(x) =  C_t(x\vert x_R)
\label{siegmund}
\end{eqnarray}

This type of property is known under the name of Siegmund duality between the two models
\cite{siegmund,Cox1983,cliff,Dette,siegmund_Intertwining,Kolo,siegmund_pathwise,Lorek,Zhao,c_DualitySpectral} with the recent generalization to various active models \cite{siegmund_runtumble,siegmund_bridge},
but has also been described otherwise in various contexts 
\cite{levy,Ciesielski-Taylor,biane,toth,tourigny}.
The Siegmund duality is a special case of the more general notion of Markov dualities
(see the reviews \cite{ReviewMohle,ReviewDuality,AlgebraicReview} with various scopes and references therein)
that have attracted a lot of interest recently
\cite{tailleurMapping,giardina_particle,giardina_transport,Redig_genetic,DualityEigen,DualityHidden}.

Note that the Siegmund duality and other Markov dualities are usually formulated 
as an equality between averaged-values of observables computed in two models.
This formulation has the advantage of producing concrete results for observables like Eq. \ref{siegmund},
but has the drawback of giving the impression that "finding dual processes is something of a black art"
as quoted in the introduction of the review \cite{ReviewDuality}.
As a consequence, the supersymmetric perspective described above
is useful to reformulate the Siegmund duality 
as an identity at the level of operators between the supersymmetric partner 
${\hat {\cal F} } = - {\cal J}  \frac{\partial}{\partial x}$ of one model
and the adjoint operator ${\mathring {\cal F}}^{\dagger} ={\mathring {\cal J} }^{\dagger} \frac{\partial}{\partial x}  $
of the dual model. Another advantage is that the duality between boundary-driven B.C. and equilibrium B.C.
described in \cite{tailleurMapping,c_boundarydriven} actually involves exactly the same 
transformations between the forces in Eq. \ref{dualityringforce} or the potentials in Eq. \ref{Uxring}.


\subsection{ Identification of the two spectral decompositions using appropriate boundary conditions}

We have already discussed the spectral decomposition of the supersymmetric partner ${\hat {\cal F} } $
in Eq. \ref{spectralPartner}  
\begin{eqnarray}
\langle x \vert e^{ t {\hat {\cal F} }} \vert x_0 \rangle 
&&=  \sum_{n=1}^{+\infty} e^{- t E_n}  j_n(x) i_n(x_0)
  \label{spectralPartnerbis}
\end{eqnarray}
with the corresponding boundary conditions for the right eigenvectors $j_n(x)$ in Eq. \ref{jnzeroagain}
and the left eigenvectors $i_n(x)$ in Eq. \ref{Oavdynfulllnbci}.
So if one wishes to push the correspondence 
${\hat {\cal F} } =  {\mathring {\cal F}}^{\dagger}   $ towards the correspondence
between the propagator of Eq. \ref{spectralPartnerbis}
and the propagator associated to ${\hat {\cal F} }^{\dagger}$
\begin{eqnarray}
 \langle x \vert  e^{t {\mathring {\cal F}}^{\dagger}  }\vert x_0 \rangle
&& =  \sum_{n=1}^{+\infty} e^{- t E_n}  {\mathring l_n }(x) {\mathring r_n }(x_0) 
\label{FPspectralring}
\end{eqnarray}
then the identification between eigenvectors yields that
\begin{eqnarray}
  {\mathring l_n }(x) =  j_n(x)
\label{mathringln}
\end{eqnarray}
satisfies the eigenvalue equation 
\begin{eqnarray}
 - E_n  {\mathring l_n }(x) = {\cal F}^{\dagger}{\mathring l_n }(x)
\label{mathringlneigen}
\end{eqnarray}
with the boundary conditions translated from Eq. \ref{jnzeroagain}
\begin{eqnarray}
  {\mathring l_n }(x) =  j_n(x)  =0 \text{ for $x=x_L$ and for $x=x_R$}
\label{mathringlnbc}
\end{eqnarray}
and that
\begin{eqnarray}
  {\mathring r_n }(x) =  i_n(x)
\label{mathringrn}
\end{eqnarray}
satisfies the eigenvalue equation 
\begin{eqnarray}
 - E_n  {\mathring r_n }(x) = {\cal F}{\mathring r_n }(x)
\label{mathringrneigen}
\end{eqnarray}
with the boundary conditions translated from Eq. \ref{Oavdynfulllnbci}
\begin{eqnarray}
   P_{eq}(x)  D(x) {\mathring r_n }(x) =0 \text{ for $x=x_L$ and for $x=x_R$}
 \label{mathringrnbc}
\end{eqnarray}

\subsection{ Discussion } 

In this section, we have shown how the reinterpretation of the supersymmetric partner ${\hat {\cal F} } = - {\cal J}  \frac{\partial}{\partial x}$ as the adjoint ${\mathring {\cal F}}^{\dagger} ={\mathring {\cal J} }^{\dagger} \frac{\partial}{\partial x}  $ of  the Fokker-Planck generator ${\mathring {\cal F}}=-  \frac{\partial}{\partial x}  {\mathring {\cal J} }$ associated to the dual force ${\mathring F}(x) =- F(x) -  D'(x)$ is useful to unify and reformulate various known Markov dualities at the level of the operator identity ${\hat {\cal F} } ={\mathring {\cal F}}^{\dagger} $.
The spectral analysis of more general Markov dualities between generators living in possibly different spaces
 is discussed in detail in the recent work \cite{c_DualitySpectral}.

Note that the duality relation ${\mathring F}(x) =- F(x) -  D'(x)$ between forces is involutive,
while in the field of supersymmetric quantum mechanics (see the reviews \cite{review_susyquantum,Mielnik_review,sasaki_reviewSusyQM,Review_factorization}),
the supersymmetric partnership is usually leveraged in order  
to construct all the eigenstates and eigenvalues via an iterative procedure
that produces a new model at each step.
In the next section, we explain how this iterative construction is based
on a different reinterpretation of the supersymmetric partner ${\hat {\cal F} }$.


\section{ Interpreting the supersymmetric partner ${\hat {\cal F} }  = - {\cal J}  \frac{\partial}{\partial x}$
as the non-conserved Fokker-Planck generator ${\tilde {\cal F}}_{nc} =  -\frac{\partial}{\partial x} {\tilde {\cal J}} - {\tilde K }(x)$ 
involving the force $  {\tilde F }(x) $ and the killing rate ${\tilde K }(x)  $}

\label{sec_nckilling}

In this section, we describe how the supersymmetric partner ${\hat {\cal F} } = - {\cal J}  \frac{\partial}{\partial x}$
 can be interpreted as the non-conserved Fokker-Planck generator ${\tilde {\cal F}}_{nc} =  -\frac{\partial}{\partial x} {\tilde {\cal J}} - {\tilde K }(x)$ involving the force $  {\tilde F }(x) $ and the killing rate ${\tilde K }(x)  $, while the diffusion coefficient $D(x)$ remains the same.

\subsection{ Correspondence between ${\hat {\cal F} } \equiv - {\cal J}  \frac{\partial}{\partial x} $
and ${\tilde {\cal F}}_{nc} \equiv  -\frac{\partial}{\partial x} {\tilde {\cal J}} - {\tilde K }(x)    $}

The supersymmetric partner ${\hat {\cal F} } =- {\cal J}  \frac{\partial}{\partial x}$ of Eq. \ref{susypartnerFP}
can also be rewritten
as the non-conserved Fokker-Planck generator ${\tilde {\cal F}}_{nc}$ 
with the same diffusion coefficient diffusion $D(x)$,
with another force $  {\tilde F }(x) $,
and with some killing rate $  {\tilde K }(x)  $ at position $x$ (the name 'killing rate' is appropriate for ${\tilde K }(x) >0$,
while ${\tilde K }(x) <0$ should be instead interpreted as the reproducing rate $(- {\tilde K }(x) >0$)
\begin{eqnarray}
{\tilde {\cal F}}_{nc} &&  \equiv - {\tilde K }(x) -\frac{\partial}{\partial x} \left(  {\tilde F }(x)   - D(x) \frac{\partial}{\partial x}  \right)
\nonumber \\
&& =\left( - {\tilde K }(x) - {\tilde F }'(x) \right) - \left( {\tilde F }(x) -D'(x) \right) \frac{\partial}{\partial x}   
  +  D(x) \frac{\partial^2}{\partial x^2} 
\label{susypartnerFPrew}
\end{eqnarray}
where the identification with ${\hat {\cal F} }  $ of Eq. \ref{susypartnerFP} leads to
\begin{eqnarray}
0 && =- {\tilde K }(x) - {\tilde F }'(x)
\nonumber \\
 F(x) && ={\tilde F }(x) -D'(x)
\label{susynewfacteurinter}
\end{eqnarray}
i.e. the identification ${\hat {\cal F} } = {\tilde {\cal F}}_{nc} $
leads to
 the parameters 
\begin{eqnarray}
{\tilde F }(x) && = F(x) +D'(x)
\nonumber \\
{\tilde K }(x) && =- {\tilde F }'(x) = -F'(x)-D''(x)
\label{susynewfacteur}
\end{eqnarray}

This identification can be equivalently found by comparing
the adjoint operator ${\hat {\cal F}}^{\dagger} $ that
involves the adjoint ${\cal J}^{\dagger} $  of Eq. \ref{defJop}
\begin{eqnarray}
 {\hat {\cal F}}^{\dagger} && = \frac{\partial}{\partial x}  {\cal J}^{\dagger} 
=     \frac{\partial}{\partial x} \left( F(x)   + \frac{\partial}{\partial x}   D(x) \right)
\nonumber \\
&& = \bigg( F'(x) +D''(x) \bigg) + \bigg( F(x)  +2 D'(x) \bigg) \frac{\partial}{\partial x} + D(x)   \frac{\partial^2}{\partial x^2}
\label{adjointIto}
\end{eqnarray}
and the adjoint of Eq. \ref{susypartnerFPrew}
\begin{eqnarray}
{\tilde {\cal F}}^{\dagger}_{nc}  = - {\tilde K }(x) + \left({\tilde F } + D'(x) \right) \frac{\partial}{\partial x} + D(x)   \frac{\partial^2}{\partial x^2}
\label{susypartnerFPrewadjoint}
\end{eqnarray}


\subsection{ Identification between the two spectral decompositions using appropriate boundary conditions}

We have already discussed the spectral decomposition of the supersymmetric partner ${\hat {\cal F} } $
in Eq. \ref{spectralPartner}  
\begin{eqnarray}
\langle x \vert e^{ t {\hat {\cal F} }} \vert x_0 \rangle 
&&=  \sum_{n=1}^{+\infty} e^{- t E_n}  j_n(x) i_n(x_0)
  \label{spectralPartnerter}
\end{eqnarray}
with the corresponding boundary conditions for the right eigenvectors $j_n(x)$ in Eq. \ref{jnzeroagain}
and the left eigenvectors $i_n(x)$ in Eq. \ref{Oavdynfulllnbci}.

So if one wishes to push the correspondence 
${\hat {\cal F} } =  {\tilde {\cal F}}_{nc}   $ towards the correspondence
between the propagator of Eq. \ref{spectralPartnerter}
and the propagator associated to $ {\tilde F}_{nc}$
\begin{eqnarray}
 \langle x \vert  e^{t {\tilde F}_{nc} }\vert x_0 \rangle
&& =  \sum_{n=1}^{+\infty} e^{- t E_n}  {\tilde r_n }(x) {\tilde l_n }(x_0) 
\label{FPspectralnc}
\end{eqnarray}
then the identification between eigenvectors yields that
\begin{eqnarray}
 {\tilde r_n }(x) =j_n(x)
\label{tildern}
\end{eqnarray}
satisfies the eigenvalue equation
\begin{eqnarray}
{\tilde F}_{nc} {\tilde r_n }(x) = -E_n {\tilde r_n }(x)
\label{tilderneigen}
\end{eqnarray}
and the boundary conditions inherited from Eq. \ref{jnzeroagain}
\begin{eqnarray}
{\tilde r_n }(x)   =0 \text{ for $x=x_L$ and for $x=x_R$}
  \label{jnzeroagaintilde}
\end{eqnarray}
and that
\begin{eqnarray}
 {\tilde l_n }(x) =i_n(x)
\label{tildeln}
\end{eqnarray}
satisfies the eigenvalue equation
\begin{eqnarray}
{\tilde F}^{\dagger}_{nc} {\tilde l_n }(x) = -E_n {\tilde l_n }(x)
\label{tildelneigen}
\end{eqnarray}
and the boundary conditions inherited from Eq. \ref{Oavdynfulllnbci}
\begin{eqnarray}
 P_{eq}(x)  D(x)  {\tilde l_n }(x) = 0  \text{ for $x=x_L$ and for $x=x_R$}
 \label{Oavdynfulllnbcitilde}
\end{eqnarray}


\subsection{ Link with the exact-solvability of Pearson diffusions : shape-invariance via supersymmetric partnership}

\label{sec_pearson}

The Pearson family of ergodic diffusions (see  \cite{pearson1895,pearson_wong,diaconis,autocorrelation,pearson_class,pearson2012,PearsonHeavyTailed,pearson2018,c_pearson}
and references therein) 
 is characterized by
a linear force
\begin{eqnarray}
F(x)  = \lambda-\gamma x
\label{pearsonF}
\end{eqnarray}
while the diffusion coefficient $D(x)$ is quadratic and vanishes at the boundaries $x_L$ and $x_R$ if these boundaries are finite
\begin{eqnarray}
D(x)  = a x^2+b x+c >0 \text { for $x \in ]x_L,x_R[$ with  $D(x_L)=0=D(x_R)$}
\label{pearsonD}
\end{eqnarray}

The correspondence between the supersymmetric partner ${\hat {\cal F} }={\tilde {\cal F}}_{nc}$
and the non-conserved generator ${\tilde {\cal F}}_{nc} $ of Eq. \ref{susypartnerFPrew}
involves the parameters of Eq. \ref{susynewfacteur}
\begin{eqnarray}
{\tilde F }(x) && = F(x) +D'(x) =  (\lambda+b) -(\gamma-2a) x \equiv {\tilde \lambda} - {\tilde \gamma } x
\nonumber \\
{\tilde K }(x) && =- {\tilde F }'(x) =- F'(x)-D''(x) = (\gamma-2a) \equiv  {\tilde \gamma }
\label{susynewfacteurbis}
\end{eqnarray}
So the new force ${\tilde F }(x) $ is linear with the following modified parameters 
with respect to the linear force $F(x)$ of Eq. \ref{pearsonF}
\begin{eqnarray}
 {\tilde \lambda} && \equiv \lambda+b
 \nonumber \\
  {\tilde \gamma } && \equiv \gamma-2a
\label{mappingPearsonF}
\end{eqnarray}
while the killing rate ${\tilde K }(x) = {\tilde \gamma }$ does not depend on $x$.
One thus obtains that the supersymmetric partner ${\hat {\cal F} }^{[\lambda,\gamma] } $
governing the dynamics of the current in the Pearson model of parameter $[\lambda,\gamma] $
can be rewritten as
 \begin{eqnarray}
{\hat {\cal F} }^{[\lambda,\gamma] } = {\tilde {\cal F}}_{nc}
= {\tilde \gamma } + {\cal F}^{[{\tilde \lambda},{\tilde \gamma}] }
\label{mappingPearson}
\end{eqnarray}
where ${\cal F}^{[{\tilde \lambda},{\tilde \gamma}] } $ is the Fokker-Planck generator 
of the Pearson model with the modified parameter $[{\tilde \lambda},{\tilde \gamma}] $.
This property for the generators is called 'shape-invariance' in the field of supersymmetric quantum mechanics 
(see the reviews \cite{review_susyquantum,Mielnik_review,sasaki_reviewSusyQM,Review_factorization}
with various scopes and references therein).

Besides this bulk property, the Pearson diffusions enjoy very specific boundary properties
as a consequence of the vanishing of the diffusion coefficient $D(x)$ at the two boundaries $x=x_L$ and $X=x_R$
in Eq. \ref{pearsonD} and in particular, the vanishing-current B.C. are automatically satisfied 
(see \cite {c_pearson} for detailed discussions).
In the present context, this means that the Pearson Fokker-Planck generator ${\cal F}^{[{\tilde \lambda},{\tilde \gamma}] } $ 
with the modified parameter $[{\tilde \lambda},{\tilde \gamma}] $
has a vanishing eigenvalue $E_0^{[{\tilde \lambda},{\tilde \gamma}] } =0$
associated to the trivial left eigenvector $l_0^{[{\tilde \lambda},{\tilde \gamma}] } (x)=1$
\begin{eqnarray}
E_0^{[{\tilde \lambda},{\tilde \gamma}] } =0
\nonumber \\
l_0^{[{\tilde \lambda},{\tilde \gamma}] } (x)=1
\label{E00Pearson}
\end{eqnarray}
As a consequence, the constant killing rate ${\tilde K }(x) =  {\tilde \gamma } $ that appears in Eq. \ref{mappingPearson}
directly represents the first-excited eigenvalue $E_1^{[\lambda,\gamma]}$ 
of the initial Fokker-Planck generator ${\cal F}^{[\lambda,\gamma]}$
\begin{eqnarray}
E_1^{[\lambda,\gamma]} =  {\tilde \gamma } = \gamma-2a
\label{E1Pearson}
\end{eqnarray}
while the corresponding left eigenvector $i_1^{[\lambda,\gamma] }(x)$ of the supersymmetric partner 
$ {\hat {\cal F} }^{[\lambda,\gamma] }$ of Eq. \ref{mappingPearson}
coincides with $l_0^{[{\tilde \lambda},{\tilde \gamma}] } (x)=1 $ of Eq. \ref{E00Pearson}
\begin{eqnarray}
i_1^{[\lambda,\gamma]} (x)= l_0^{[{\tilde \lambda},{\tilde \gamma}] } (x)=1
\label{i1Pearson}
\end{eqnarray}
The corresponding relations of Eq. \ref{splittingrl} for $n=1$ 
are then useful to obtain that the first-exited left eigenvector $l_1^{[\lambda,\gamma]} (x) $ of the initial Pearson 
Fokker-Planck generator
${\cal F}^{[\lambda,\gamma] } $
\begin{eqnarray}
 l_1^{[\lambda,\gamma]} (x)   && = {\cal J}^{\dagger}    i_1^{[\lambda,\gamma]} (x) 
 = \bigg(  F(x)+D'(x) + D(x) \frac{\partial}{\partial x} \bigg ) 1= {\tilde F }(x) = {\tilde \lambda} - {\tilde \gamma } x
\nonumber \\
- \frac{\partial}{\partial x}  l_1^{[\lambda,\gamma]} (x) ( x)  && =  E_1^{[\lambda,\gamma]}    i_1^{[\lambda,\gamma]} ( x)  
 = {\tilde \gamma }
\label{splittingrlpearson}
\end{eqnarray}
 coincides with the linear force ${\tilde F }(x) $ and is thus a polynomial of degree one.
Via iteration, one recovers that all the excited eigenvalues $E_n^{[\lambda,\gamma]} $
of the initial Pearson Fokker-Planck generator ${\cal F}^{[\lambda,\gamma] } $
can be explicitly computed with their corresponding eigenvectors $l_n^{[\lambda,\gamma]}$
that reduce to polynomials of degree $n$
(see \cite {c_pearson} for more detailed discussions and for the various types of Pearson diffusions depending on whether
$x_L$ and $x_R$ are finite or infinite).

\subsection{ Discussion } 

In this section, we have explained how the interpretation of the supersymmetric partner 
${\hat {\cal F} }={\tilde {\cal F}}_{nc}$
as the non-conserved generator ${\tilde {\cal F}}_{nc} $ involving the killing rate $ {\tilde K }(x)$
is useful to make the link with the standard use of supersymmetric partnership
in the field of supersymmetric quantum mechanics (see the reviews \cite{review_susyquantum,Mielnik_review,sasaki_reviewSusyQM,Review_factorization}),
where the goal is to construct all the eigenstates and eigenvalues via an iterative procedure,
and to understand why the Pearson diffusions are exactly solvable with the shape-invariance property of
Eq. \ref{mappingPearson}
that involves a constant killing rate.
 

\section{ Conclusions}

\label{sec_conclusion}

 In the main text, we have first explained why it is useful to analyze 
 one-dimensional reversible diffusion processes 
 involving arbitrary forces $F(x)$ and arbitrary diffusion coefficients $D(x)$
 by considering the probabilities $P_t(x)$ and the currents $J_t(x)$ on the same footing.
 The Fokker-Planck generator governing the dynamics of the probability $P_t(x)$ alone
 is then naturally factorized
 ${\cal F}  =  -  \frac{\partial}{\partial x} {\cal J}$ in terms of two first-order differential operators 
 coming from the continuity equation, while the supersymmetric partner ${\hat {\cal F} }=  - {\cal J}  \frac{\partial}{\partial x}$ directly governs the dynamics of the current $J_t(x)$ alone and has thus a very clear physical meaning.
 We have explained the links with the standard mapping towards some hermitian quantum Hamiltonian $H=Q^{\dagger} Q$  via a similarity transformation, and with 
the factorization of the adjoint operator $ {\cal F}^{\dagger}={\cal J}^{\dagger} \frac{\partial}{\partial x}$ 
that is standard in the mathematical literature
 $ {\cal F}^{\dagger} =\frac{d}{d m(x) } \frac{d}{d s(x) } $ in terms of the scale function $s(x)$ and speed measure $m(x)$.

After these physically-motivated unifying reformulations of various previously known results 
of the physical and mathematical literatures
in sections \ref{sec_diffusionPJ} and  \ref{sec_diffusionP}, 
we have turned to the more novel part of the present work
where the goal was to analyze the properties of the 
supersymmetric partner ${\hat {\cal F} }=  - {\cal J}  \frac{\partial}{\partial x}$ 
governing the dynamics of the current $J_t(x)$:

$\bullet$ In section \ref{sec_diffusionJ}, we have explained how the spectral properties of 
the supersymmetric partners ${\cal F} $ and ${\hat {\cal F} } $
are directly related via the two intertwining relations ${\cal J}{\cal F}   =  - {\cal J} \frac{\partial}{\partial x} {\cal J} =  {\hat {\cal F} } {\cal J}$ and ${\cal F} \frac{\partial}{\partial x}  =  -  \frac{\partial}{\partial x} {\cal J} \frac{\partial}{\partial x} = \frac{\partial}{\partial x}{\hat {\cal F} }  $ between ${\cal F} $ and ${\hat {\cal F} } $ : 
their non-vanishing eigenvalues $E_n \ne 0$ thus coincide, while their respective 
corresponding right and left eigenvectors are related via the applications of 
the first-order differential operators $\frac{\partial}{\partial x} $ and ${\cal J} $.
These properties can be thus considered as the non-hermitian generalization 
of the standard method used in the field of
supersymmetric hermitian quantum Hamiltonians.

$\bullet$ In order to better understand the differences between 
the Markov generator  ${\cal F}  =  -  \frac{\partial}{\partial x} {\cal J}$
governing governing the dynamics of the probability $P_t(x)$ 
and  its supersymmetric partner ${\hat {\cal F} }=  - {\cal J}  \frac{\partial}{\partial x}$ governing the dynamics of the current $J_t(x)$ that are a priori very different, since
 the probability $P_t(x)$ is positive, normalized and converges towards 
a steady state $P_{eq}(x)$, 
while the current $J_t(x)$ can be either positive or negative, is not normalized and converges towards zero $J_{eq}=0$, we have analyzed two very different re-interpretations of the supersymmetric partner  ${\hat {\cal F} } = - {\cal J}  \frac{\partial}{\partial x}$ :
  
 (i) In Section \ref{sec_duality},
 we have reinterpreted ${\hat {\cal F} } $ as the adjoint ${\mathring {\cal F}}^{\dagger} ={\mathring {\cal J} }^{\dagger} \frac{\partial}{\partial x}  $ of  the Fokker-Planck generator ${\mathring {\cal F}}=-  \frac{\partial}{\partial x}  {\mathring {\cal J} }$ associated to the dual force ${\mathring F}(x) =- F(x) -  D'(x)$. We have explained why this 
 interpretation is useful to unify and reformulate various known Markov dualities at the level of the operator identity ${\hat {\cal F} } ={\mathring {\cal F}}^{\dagger} $, instead of the standard formulation of Markov dualities 
 via an equality between averaged-values of observables computed in two models.
As discussed in detail in the recent work \cite{c_DualitySpectral},
 the spectral reformulation of Markov dualities 
 is also possible and very useful for generators living in different spaces.

 (ii) In section \ref{sec_nckilling}, we have analyzed an alternative reinterpretation of ${\hat {\cal F} } $ as the non-conserved Fokker-Planck generator ${\tilde {\cal F}}_{nc} =  -\frac{\partial}{\partial x} {\tilde {\cal J}} - {\tilde K }(x)$  involving the force $  {\tilde F }(x) =F(x) +D'(x)$ and the killing rate ${\tilde K }(x)=-F'(x)-D''(x)  $.
 This reinterpretation 
  is useful to make the link with the notion of shape-invariance of the field of supersymmetric quantum mechanics and to
 recover why Pearson diffusions involving a linear force $F(x)$ and a quadratic diffusion coefficient $D(x)$ are exactly solvable via their shape-invariance property involving constant killing rates.

 Our main conclusion is thus 
that the present supersymmetric perspective 
clarifies the spectral properties of the generators governing
the dynamics of the probabilities $P_t(x)$ and of the currents $J_t(x)$, 
unifies various notions of Markov dualities at the level of operator identities,
and directly leads to the identification of models with shape-invariance-exact-solvability.
In the various Appendices, we describe how all these physical ideas can be also applied to reversible Markov jump processes on the one-dimensional lattice with arbitrary nearest-neighbors transition rates $w(x \pm 1,x)$, via the replacement of derivatives by finite differences. The similarities and the differences between Markov models defined in continuous space and discrete space are actually useful to better understand both.

As recalled in the Introduction, this point of view is also useful to make the link
with other recent works concerning non-equilibrium Markov processes,
either in dimension $d=1$ with other boundary conditions leading to non-equilibrium steady currents,
as described in [14,18] for periodic boundary conditions and Boundary-driven-by-reservoirs respectively,
or in arbitrary dimensions as discussed in 
\cite{c_susySVD} both for Fokker-Planck generators in dimension $d>1$
(where the current $\vec J_t(\vec x)$ is a d-dimensional vector), 
and for Markov jump generators on arbitrary graphs.
Finally, let us mention 
that all these ideas concerning a single one-dimensional Markov process
are also useful in the field of integrable models of interacting Markov processes 
(see \cite{TheoExchangeSiegmund,ExchangemsTheo,TheoExactSolv1,TheoExactSolv2}
and references therein).


\appendix

\section{ Reversible Markov-jump dynamics in $d=1$ in terms of probabilities $P_t(y)$ and currents $J_t(x)$}

\label{app_JumpPJ}

In this Appendix, we describe how the section \ref{sec_diffusionPJ} of the main text
can be adapted to Markov-jump dynamics with nearest-neighbor transition rates $w(x \pm 1,x)$ 
on the one-dimensional lattice $0 \leq x \leq N$.

 \subsection{ Continuity equation for the probability $P_t(x)$ involving the currents $J_t(x) $   }

The probability $P_t(y)$ is defined on the $(N+1)$ sites $y=0,1,..,N$,
while the current $J_t(x) $ is defined on the $N$ links $(x,x+1)$ with $x=0,1,..,N-1$
(note than another possibility is to label the currents $J_t\left( x+\frac{1}{2} \right)$ via the middle-point $\left( x+\frac{1}{2} \right)$ of the associated link $(x,x+1)$ as described in \cite{c_boundarydriven})
\begin{eqnarray}
J_t(x) \equiv w(x+1,x) P_t(x) - w(x,x+1) P_t(x+1) \equiv \sum_{y=0}^N {\bold J}(x,y) P_t(y)
\label{jlink}
\end{eqnarray}
and is computed from the probabilities $P_t(y) $
via the application of the current matrix ${\bold J}(x,y)$ of size $N \times (N+1)$ 
with matrix elements only on the diagonal $x=y $ and on the upper-diagonal $x+1=y$
\begin{eqnarray}
 {\bold J}(x,y) = w(x+1,x) \delta_{x,y}  - w(x,x+1) \delta_{x+1,y}
\label{currentMatrix}
\end{eqnarray}
This matrix ${\bold J} $ is the analog of the current differential operator ${\cal J}$ of Eq. \ref{defJop} of the main text.

When the rates are unity in the current matrix of Eq. \ref{currentMatrix},
one obtains the matrix of size $N \times (N+1)$
with matrix elements only on the diagonal $x=y $ and on the upper-diagonal $x+1=y$
\begin{eqnarray}
 {\bold I}(x,y) = \delta_{x,y}  -  \delta_{x+1,y}
\label{IncidenceMatrix}
\end{eqnarray}
that can be applied to the probability $P_t(y)$ (or any other function defined on sites) to compute the discrete difference
\begin{eqnarray}
 \sum_{y=0}^N {\bold I}(x,y) P_t(y) = P_t(x)  -  P_t(x+1)
\label{IncidenceMatrixapplied}
\end{eqnarray}
while its adjoint of size $(N+1) \times N$
with matrix elements only on the diagonal $y=x $ and on the lower-diagonal $y=x+1$
\begin{eqnarray}
 {\bold I}^{\dagger}(y,x) =  {\bold I}(x,y)  = \delta_{y,x}  -  \delta_{y,x+1}
\label{IncidenceMatrixTransposed}
\end{eqnarray}
can be applied to the current $J_t(x)$ (or any other function defined on links) to compute the discrete difference
\begin{eqnarray}
\sum_{x=0}^{N-1} {\bold I}^{\dagger}(y,x) J_t(x)=  J_t(y)  -  J_t(y-1)
\label{IncidenceMatrixTransposedapplied}
\end{eqnarray}
The matrices ${\bold I} $ and ${\bold I}^{\dagger} $ are the analogs of the derivative operators 
$\left( - \frac{\partial}{\partial x}\right)$ and $\left(  \frac{\partial}{\partial y}\right)$that appear in the main text

The continuity equation describing the evolution of the probability $P_t(y)$
involves the difference of the currents of Eq. \ref{IncidenceMatrixTransposedapplied}
and can be thus rewritten as the application of the matrix ${\bold I}^{\dagger} $ to the current $J_t(x)$
\begin{eqnarray}
 \partial_t P_t(y)   && =  J_t(y-1) -  J_t(y) = - \sum_{x=0}^{N-1}   {\bold I}^{\dagger}(y,x) J_t(x)
 \label{continuityDiscrete}
\end{eqnarray}

 \subsection{ Vanishing-current Boundary Conditions $J_t(-1) =0 = J_t(N)$   }
 
 The analog of the vanishing-current Boundary Conditions of Eq. \ref{vanishinhJ} read
\begin{eqnarray}
J_t(-1) && =0
 \nonumber \\
J_t(N) && =0
\label{vanishJDiscrete}
\end{eqnarray}
so that the dynamics at the two boundary-sites $y=0$ and $y=N$ reduce to
\begin{eqnarray}
 \partial_t P_t(0) && = -  J_t(0)
 \nonumber \\
 \partial_t P_t(N)   && =  J_t(N-1)
\label{continuityDiscreteBC}
\end{eqnarray}

Other B.C. like Periodic or Boundary-driven leading to non-equilibrium have been also much studied
in the discrete case (see \cite{c_ring,c_boundarydriven} and references therein).


\section{ Dynamics for the probability $P_t(x)$ governed by the factorized Markov matrix 
${\bold M} = - {\bold I}^{\dagger} {\bold J}$   }

\label{app_JumpP}

In this Appendix, we describe how the section \ref{sec_diffusionP} of the main text
can be adapted to the Markov-jump dynamics.


\subsection{ Factorized Markov matrix 
${\bold M} = - {\bold I}^{\dagger} {\bold J}$ governing the Master Equation for the probability $P_t(y)$}

Plugging Eq. \ref{jlink} into Eq \ref{continuityDiscrete} leads to the standard master equation
\begin{eqnarray}
 \partial_t P_t(x)   && =  w(x,x-1) P_t(x-1)  + w(x,x+1) P_t(x+1) - \bigg[ w(x-1,x) +w(x+1,x) \bigg]P_t(x)
 \nonumber \\
 && = \sum_{y=0}^N {\bold M}(x,y) P_t(y)
\label{master}
\end{eqnarray}
where the factorized Markov matrix matrix ${\bold M}=- {\bold I}^{\dagger} {\bold J}$ 
is tridiagonal with off-diagonal elements given by positive transition rates $w(.,.)$
\begin{eqnarray}
{\bold M}(x,x-1) && =  w(x,x-1)  
 \nonumber \\
{\bold M}(x,x-1) && =w(x,x+1)
\label{masteroff}
\end{eqnarray}
while the diagonal element $ {\bold M}(x,x) $ is the opposite of the total rate out of site $x$ 
\begin{eqnarray}
 {\bold M}(x,x) = - \bigg[ w(x-1,x) +w(x+1,x) \bigg] = - \sum_{x' \ne x} {\bold M}(x',x) 
\label{masterdiag}
\end{eqnarray}
as it should to conserve the total probability.

To make the link with the notations of Eq. \ref{Ux} of the main text, it is useful to introduce the following parametrization of the transition rates 
\begin{eqnarray}
 w(x+1,x)  =D(x) e^{ \frac{ U(x)-U(x+1) }{2} }
 \nonumber \\
  w(x,x+1)  =D(x) e^{  \frac{ U(x+1)-U(x) }{2} }
\label{rates}
\end{eqnarray}
where the diffusion coefficient $D(x)$ defined on the links $(x,x+1)$ with $x=0,..,N-1$
and the potential $U(y)$ defined on the sites $y=0,..,N$ can be computed from the transition rates via
\begin{eqnarray}
D(x) && = \sqrt{ w(x+1,x)   w(x,x+1) }
\nonumber \\
U(x+1)-U(x) && = \ln \left( \frac{  w(x,x+1) }{  w(x+1,x) } \right)
\label{ratesInversion}
\end{eqnarray}

Then the current matrix ${\bold J}(x,y)$ of Eq. \ref{currentMatrix} can be rewritten in the factorized form
involving the matrix ${\bold I}$ of Eq. \ref{IncidenceMatrix}
\begin{eqnarray}
 {\bold J}(x,y) && = D(x) e^{ \frac{ U(x)-U(x+1) }{2} } \delta_{x,y}  - D(x) e^{  \frac{ U(x+1)-U(x) }{2} } \delta_{x+1,y}
 \nonumber \\
 && = D(x) e^{-  \frac{ U(x)+U(x+1) }{2} } {\bold I}(x,y) e^{U (y)}
\label{currentMatrixUD}
\end{eqnarray}
that is the analog of Eq. \ref{defJopU} of the main text,
so that the current $J_t(x) $ on the link $(x,x+1)$ of Eq. \ref{jlinkUD}
becomes
\begin{eqnarray}
J_t(x) = D(x) e^{ - \frac{ U(x)+U(x+1) }{2} } \left[ e^{U(x)} P_t(x) -  e^{  U(x+1) } P_t(x+1) \right]
\label{jlinkUD}
\end{eqnarray}

The corresponding form of the Markov matrix ${\bold M}=- {\bold I}^{\dagger} {\bold J}$
\begin{eqnarray}
{\bold M}(z,y) && =- \sum_{x=0}^{N-1} {\bold I}^{\dagger}(z,x) {\bold J}(x,y)
 \nonumber \\
 && = - \sum_{x=0}^{N-1} {\bold I}^{\dagger}(z,x) D(x) e^{-  \frac{ U(x)+U(x+1) }{2} } {\bold I}(x,y) e^{U (y)}
\label{MatrixMUD}
\end{eqnarray}
is the analog of Eq. \ref{defJopU}.


\subsection{ The two independent explicit solutions of 
$0 = {\bold M} P_*(y) $ in the bulk }

The analog of Eq. \ref{rhoeq} is 
 the equilibrium in the potential $U(y)$
\begin{eqnarray}
  \rho_{eq}(y) =  e^{  -U(y) } 
\label{steadyeqjump}
\end{eqnarray}
while the analog of Eq. \ref{rhoneq} reads 
\begin{eqnarray}
  \rho^{noneq}_*(y) && = e^{-U(y) }  \sum_{z=0}^{y-1} \frac{ e^{\frac{ U(z)+U(z+1) }{2} }  } { D(z) } 
\label{steadynoneqjump}
\end{eqnarray}


\subsection{ Factorized adjoint matrix 
${\bold M}^{\dagger}  = - {\bold J}^{\dagger} {\bold I} $ governing the dynamics of observables $O(y)$ of the sites}

The average of an observable $O(y)$ computed with the probability $P_t(y)$ is the analog of Eq. \ref{Oav}
\begin{eqnarray}
 \sum_{y=0}^N O(y) P_t(y)  = \langle O \vert P_t \rangle
 \label{Oavdiscrete}
\end{eqnarray}
Its dynamics can be analyzed using the master Equation \ref{master}
\begin{eqnarray}
 \partial_t \left[ \sum_{y=0}^N O(y) P_t(y) \right] &&  =   \sum_{y=0}^N O(y) \left[ \partial_t P_t(y) \right]
 =\sum_{y=0}^N O(y) \sum_{z=0}^N {\bold M}(y,z) P_t(z)
 \nonumber \\
 && =   \sum_{z=0}^N  P_t(z) \bigg[ \sum_{y=0}^N O(y) {\bold M}(y,z) \bigg]
 = \sum_{z=0}^N  P_t(z) \bigg[ \sum_{y=0}^N  {\bold M}^{\dagger} (z,y) O(y) \bigg] =  \langle {\bold M}^{\dagger} O \vert P_t \rangle
\label{masterobser}
\end{eqnarray}
which is the analog of Eq. \ref{Oavdynfull} when the B.C. are taken into account in the finite matrices ${\bold M} $ and ${\bold M}^{\dagger} $.

If one uses the explicit expressions of the matrix elements ${\bold M}(y,z) $ 
of Eqs \ref{masteroff}
\ref{masterdiag}
in terms of the transition rates, one obtains the familiar expression
\begin{eqnarray}
 \partial_t \left[ \sum_{y=0}^N O(y) P_t(y) \right] &&  
 = \sum_{z=0}^N  P_t(z) \bigg[  O(z+1) {\bold M}(z+1,z) + O(z-1) {\bold M}(z-1,z) + O(z) {\bold M}(z,z)\bigg]
 \nonumber \\
 && 
 =  \sum_{z=0}^N  P_t(z) \bigg[  \bigg( O(z+1)-O(z) \bigg) w(z+1,z) +\bigg( O(z-1)-O(z) \bigg) w(z-1,z) \bigg]
\label{masterobserexpli}
\end{eqnarray}

The factorized form of the current matrix of Eq. \ref{currentMatrixUD}
translates for the adjoint matrix into
\begin{eqnarray}
 {\bold J}^{\dagger}(y,x)  = {\bold J}(x,y) && = D(x) e^{-  \frac{ U(x)+U(x+1) }{2} } {\bold I}(x,y) e^{U (y)}
 \nonumber \\
 && = e^{U (y)}{\bold I}^{\dagger} (y,x)  D(x) e^{-  \frac{ U(x)+U(x+1) }{2} } 
\label{currentMatrixUDadjoint}
\end{eqnarray}
that is the analog of Eq. \ref{defJopUadjoint}.
The corresponding factorized form of the adjoint matrix ${\bold M}=- {\bold I}^{\dagger} {\bold J}$ 
\begin{eqnarray}
{\bold M}^{\dagger} (y,z) && =- \sum_{x=0}^{N-1} {\bold J}^{\dagger}(y,x) {\bold I}(x,z)
 \nonumber \\
 && = - \sum_{x=0}^{N-1} e^{U (y)}{\bold I}^{\dagger} (y,x)  D(x) e^{-  \frac{ U(x)+U(x+1) }{2} }  {\bold I}(x,z)
\label{MatrixMUDadjoint}
\end{eqnarray}


\subsection{ The two independent explicit solutions of 
$0 = {\bold M}^{\dagger} O(y) $ in the bulk }

The analog of Eq. \ref{l0unity} is also unity and associated to the conservation of probability
\begin{eqnarray}
l_0(y) && = 1 
\label{l0unitydiscrete}
\end{eqnarray}
while the analog of the scale function of Eq. \ref{Sx} reads
\begin{eqnarray}
s(y)  =  \sum_{x=0}^{y-1} \frac{ e^{\frac{ U(x)+U(x+1)}{2} }  } { D(x) } 
\ \ \ \text{ with $s(0)=0$ and the positive increments} \ \ s(y+1)-s(y) =\frac{ e^{\frac{ U(y)+U(y+1)}{2} }  } { D(y) }
\label{Sxdiscrete}
\end{eqnarray}


\subsection{ Physical interpretation of the discrete speed-measure $m(y)$
and of the discrete scale function $s(y)$ }

The cumulative equilibrium distribution $C_{eq}(x)$ analog to Eq. \ref{Ceqmx}
\begin{eqnarray}
C_{eq}(y) \equiv \sum_{z=0}^y P_{eq}(z)  =\sum_{z=0}^y \left( \frac{e^{-U(z) } }{ \sum_{y'=0}^N e^{-U(y') }}  \right) = \frac{ m(y) }
{ m(N) } \ \ \text{ with $C_{eq}(-1)=0$ and $C_{eq}(N)=1 $ }
\label{Ceqmxdiscrete}
\end{eqnarray}
can be used to defined the analog of the speed measure
\begin{eqnarray}
m(y)  \equiv \sum_{z=0}^y e^{-U(z) } 
\label{mydiscrete}
\end{eqnarray}
with its positive increment analogous to Eq. \ref{rhoeqmx}
\begin{eqnarray}
m(y)-m(y-1) = e^{- U(y)} 
\label{rhoeqmxdiscrete}
\end{eqnarray}

When the Markov jump process starts at position $y$,
the probability $E^{N}(y)$ to reach the right boundary $N$
before the left boundary $0$, 
and the complementary
probability $E^{0}(y)= 1- E^{N}(y)$ to reach the left boundary $0$ before the right boundary $N$,
 can be written in terms of the scale function $s(y)$ of Eq. \ref{Sxdiscrete}
 as in Eq. \ref{ExitRx}
\begin{eqnarray}
E^{N}(x) = \frac{ s(y) }{ s(N) } = 1- E^{0}(y)
  \label{ExitRxdiscrete}
\end{eqnarray}
Indeed, the probability $E^{N}(x) $ should be annihilated by the adjoint matrix ${\bold M}^{\dagger} $ 
 \begin{eqnarray}
0= {\bold M}^{\dagger} E^{N}(x) 
\label{adjointzeroExitdiscrete}
\end{eqnarray}
and can be thus written as a linear combination of the two independent solutions $l_0(y)=1$ and $s(y)$,
where the two constants are determined by the two boundary conditions at $y=0$ and $y=N$
 \begin{eqnarray}
 E^{N}(N) && =1
 \nonumber \\
 E^{N}(0) && =0
\label{BCExitdiscrete}
\end{eqnarray}


\subsection{ Similarity transformation between the Markov matrix $ {\bold M}$ and the
hermitian supersymmetric quantum Hamiltonian 
${\bold H}(z,y)  \equiv - e^{ \frac{U(z)}{2} }  {\bold M} (z,y) e^{- \frac{U(y)}{2} } $ }

The form of Eq. \ref{currentMatrixUD}
for the Markov matrix ${\bold M}(z,y)  $
yields that the similarity transformation towards the matrix
\begin{eqnarray}
{\bold H}(z,y)  && \equiv - e^{ \frac{U(z)}{2} }  {\bold M} (z,y) e^{- \frac{U(y)}{2} }
= e^{ \frac{U(z)}{2} } \sum_{x=0}^{N-1} {\bold I}^{\dagger}(z,x) D(x) e^{-  \frac{ U(x)+U(x+1) }{2} } 
{\bold I}(x,y)  e^{ \frac{U(y)}{2} } 
\nonumber \\
&& \equiv \sum_{x=0}^{N-1} {\bold Q}^{\dagger}(z,x)  {\bold Q}(x,y)
\label{Hsusydiscrete}
\end{eqnarray}
leads to the supersymmetric form ${\bold H}={\bold Q}^{\dagger} {\bold Q} $ involving the matrix ${\bold Q} $
and its adjoint ${\bold Q}^{\dagger}$
\begin{eqnarray}
{\bold Q}(x,y) && \equiv     \sqrt{ D(x) e^{-  \frac{ U(x)+U(x+1) }{2} } } {\bold I}(x,y)  e^{ \frac{U(y)}{2} } 
\nonumber \\
{\bold Q}^{\dagger}(z,x) && 
 =e^{ \frac{U(z)}{2} }  {\bold I}^{\dagger}(z,x) \sqrt{ D(x) e^{-  \frac{ U(x)+U(x+1) }{2} } }
\label{HsusyQdiscrete}
\end{eqnarray}
that are the analogs of the first order operators $Q$ and $Q^{\dagger}$ of Eq. \ref{HsusyQ}.


\subsection{ Spectral decomposition of the Markov propagator}

The analog of the spectral decomposition of Eq. \ref{FPspectralq}
\begin{eqnarray}
 P_t(x \vert x_0)  \equiv \langle x \vert e^{ {\bold M} t} \vert x_0 \rangle  
&&  =\sum_{n=0}^{N} e^{- t E_n}  r_n(x) l_n(x_0)
= P_{eq}(x) + \sum_{n=1}^{N} e^{- t E_n}  r_n(x) l_n(x_0)
\label{Mspectral}
\end{eqnarray}
involves the $(N+1)$ eigenvalues $(-E_n) \leq 0$ of the Markov matrix ${\bold M} $,  while
 the corresponding right eigenvectors $\vert r_n \rangle $ 
and left eigenvectors $\langle l_n \vert $ satisfy the eigenvalues equations
\begin{eqnarray}
-  E_n \vert r_n \rangle && =  {\bold M} \vert r_n \rangle
  \nonumber \\
- E_n \langle l_n \vert && =  \langle l_n \vert{\bold M}  
\label{spectralrl}
\end{eqnarray}
and form a bi-orthogonal basis with the orthonormalization and closure relations
\begin{eqnarray}
\delta_{n,n'} && = \langle l_n \vert  r_{n'} \rangle = \sum_{x=0}^N \langle l_n \vert x \rangle \langle x \vert r_{n'} \rangle
\nonumber \\
{\bold 1} && = \sum_{n=0}^{N}  \vert r_n \rangle \langle l_n \vert 
\label{orthorl}
\end{eqnarray}

The vanishing eigenvalue $E_0=0$ is associated 
to the left eigenvector unity $l_0(x)=1$ 
while the right eigenvector corresponds to the equilibrium steady state $r_0(x)=P_{eq}(x)$ 
\begin{eqnarray}
  E_0 && =0
  \nonumber \\
 l_0 (x)&& =1
  \nonumber \\
 r_0 (x)&& =P_{eq}(x)
\label{rlzero}
\end{eqnarray}

As in Eq. \ref{FPspectralqleft}, it is often simpler to rewrite the spectral decomposition of Eq. \ref{Mspectral}
\begin{eqnarray}
P_t(x \vert x_0)  = P_{eq}(x) \left[ \sum_{n=0}^{N} e^{- t E_n}  l_n(x) l_n(x_0) \right]
= P_{eq}(x) \left[ 1 + \sum_{n=1}^{N} e^{- t E_n}  l_n(x) l_n(x_0) \right]
\label{Mspectralleft}
\end{eqnarray}
in terms of the left eigenvectors $l_n(x)$ that form an orthogonal property with respect to the equilibrium measure $P_{eq}(x)$
\begin{eqnarray}
\delta_{n n' } =  \sum_{x=0}^N   l_n(x) l_{n'}(x) P_{eq}(x)
 \label{ortholrldiscrete}
\end{eqnarray}


\section{ Dynamics of the currents governed by the supersymmetric partner ${\hat {\bold M} } =  - {\bold J}^{\dagger} {\bold I}$ 
  }

\label{app_JumpJ}

In this Appendix, we describe how the section \ref{sec_diffusionJ} of the main text
can be adapted to the Markov-jump dynamics.

\subsection{ Dynamics of the currents $J_t(x)$ governed by the supersymmetric partner ${\hat {\bold M} } =  - {\bold J}^{\dagger} {\bold I}$ 
of the Markov matrix ${\bold M} = - {\bold I}^{\dagger} {\bold J}$ }

The dynamics of the $N$ currents $J_t(x)$ of Eq. \ref{jlink}
associated to the links $(x,x+1)$ with $x=0,..,N-1$
\begin{eqnarray}
\partial_t J_t(x) &&= w(x+1,x) \partial_t P_t(x) - w(x,x+1) \partial_t P_t(x+1)
\nonumber \\
&& = w(x+1,x) \left[J_t(x-1) -  J_t(x) \right]
- w(x,x+1) \left[J_t(x) -  J_t(x+1) \right]
\nonumber \\
&& = \sum_{z=0}^{N-1} {\hat {\bold M} } (x,z) J_t (z)
\label{jlinkdyn}
\end{eqnarray}
is governed by the supersymmetric partner ${\hat {\bold M} } =  - {\bold J} {\bold I}^{\dagger}$ of the Markov matrix 
${\bold M} = - {\bold I}^{\dagger} {\bold J}$ with the off-diagonal elements
\begin{eqnarray}
{\hat {\bold M} } (x,x-1) && = w(x+1,x) 
\nonumber \\
{\hat {\bold M} } (x,x+1) && = w(x,x+1) 
\label{hatwoff}
\end{eqnarray}
while the diagonal element
\begin{eqnarray}
{\hat {\bold M} } (x,x) && = - w(x+1,x) - w(x,x+1)= - {\hat {\bold M} } (x,x-1) - {\hat {\bold M} } (x,x+1) 
 = - \sum_{z \ne x} {\hat {\bold M}} (x,z)
\label{hatwdiag}
\end{eqnarray}
is the opposite of the sum of the transitions rates towards $x$.

The factorized form of Eq. \ref{currentMatrixUD} for the matrix ${\bold J}$,
leads to the factorized form 
\begin{eqnarray}
{\hat {\bold M} } (x,z) && =- \sum_{y=0}^{N} {\bold J}(x,y)  {\bold I}^{\dagger}(y,z)
 \nonumber \\
 && = -  \sum_{y=0}^{N} D(x) e^{-  \frac{ U(x)+U(x+1) }{2} } {\bold I}(x,y) e^{U (y)} {\bold I}^{\dagger}(y,z)
\label{currentMatrixUDsusy}
\end{eqnarray}
that is the analog of Eq. \ref{susypartnerFPU}.


\subsection{  Intertwining relations between the supersymmetric partners 
${\hat {\bold M} } =  - {\bold J} {\bold I}^{\dagger}$ of the Markov matrix 
${\bold M} = - {\bold I}^{\dagger} {\bold J}$ }

The Markov matrix ${\bold M} = - {\bold I}^{\dagger} {\bold J}$
and its supersymmetric partner 
${\hat {\bold M} } =  - {\bold J} {\bold I}^{\dagger}$ 
satisfy the intertwining relations
\begin{eqnarray}
{\bold J}{\bold M}   &&=  - {\bold J} {\bold I}^{\dagger} {\bold J} = {\hat {\bold M} } {\bold J}
\nonumber \\
{\bold M} {\bold I}^{\dagger}  
&&= - {\bold I}^{\dagger} {\bold J} {\bold I}^{\dagger} = {\bold I}^{\dagger}{\hat {\bold M} } 
  \label{interwJI}
\end{eqnarray}
that are the analogs of Eqs \ref{interwJ}
and \ref{interwderivative}
with similar consequences for their right and left eigenvectors.
Let us describe the case of left eigenvectors :
 the eigenvalue Eq. \ref{spectralrl} for the excited left eigenvector $ \langle l_n \vert  $ 
of ${\bold M} =- {\bold I}^{\dagger} {\bold J} $ can be split into
the two matrix equations
\begin{eqnarray}
 E_n \langle  i_n \vert && =   \langle l_n \vert {\bold I}^{\dagger}
\nonumber \\
 \langle l_n \vert  && =   \langle   i_n \vert {\bold J} 
\label{spectrallin}
\end{eqnarray}
involving the bra $\langle   i_n \vert $ which is a left eigenvector of the supersymmetric partner 
${\hat {\bold M} } \equiv -{\bold J}  {\bold I}^{\dagger} $ 
associated to the eigenvalue $(-E_n)$
\begin{eqnarray}
 E_n  \langle i_n \vert  
=  \langle l_n \vert {\bold I}^{\dagger}
=   \langle i_n \vert {\bold J}   {\bold I}^{\dagger} 
= -  \langle i_n \vert {\hat {\bold M} }
\label{excitedpartnerleft}
\end{eqnarray}

The explicit versions of Eqs \ref{spectrallin} read using 
the matrix elements of Eqs \ref{IncidenceMatrixTransposed}
and \ref{currentMatrix}
\begin{eqnarray}
 E_n i_n(x) && =   \langle l_n \vert {\bold I}^{\dagger} \vert x \rangle = \sum_y l_n(y) {\bold I}^{\dagger} (y,x) 
 = l_n(x)  - l_n(x+1)
\nonumber \\
  l_n (y) && =   \langle   i_n \vert {\bold J} \vert y \rangle = \sum_x i_n(x)  {\bold J} (x,y) 
  =   i_n(y)w(y+1,y)   - i_n(y-1) w(y-1,y) 
\label{spectrallinexpli}
\end{eqnarray}


\subsection{ Link with the supersymmetric partner ${\breve {\bold H}} = {\bold Q} {\bold Q}^{\dagger} $ 
of the quantum Hamiltonian ${\bold H}={\bold Q}^{\dagger} {\bold Q}  $}

While $ {\bold H}={\bold Q}^{\dagger} {\bold Q} $ of Eq. \ref{Hsusydiscrete} is an $(N+1)\times (N+1)$ matrix
associated to the sites,
the supersymmetric partner ${\breve {\bold H}} = {\bold Q} {\bold Q}^{\dagger} $ is an $N \times N$ matrix
associated to the links.
Its matrix elements obtained using Eq. \ref{HsusyQdiscrete} 
\begin{eqnarray}
{\breve {\bold H}} (x,x')  &&=  \sum_{y=0}^{N} {\bold Q}(x,y) {\bold Q}^{\dagger}(y,x') 
\nonumber \\
&&=  \sum_{y=0}^{N}  \sqrt{ D(x) e^{-  \frac{ U(x)+U(x+1) }{2} } } {\bold I}(x,y)  e^{ U(y) }  
{\bold I}^{\dagger}(y,x') \sqrt{ D(x') e^{-  \frac{ U(x')+U(x'+1) }{2} } }
\label{Hsusydiscretepartner}
\end{eqnarray}
is related to the supersymmetric partner ${\hat {\bold M} } (x,x') $ of Eq. \ref{currentMatrixUDsusy}
via the similarity transformation 
\begin{eqnarray}
{\breve {\bold H}} (x,x') = - \frac{1}{ \sqrt{ D(x) e^{-  \frac{ U(x)+U(x+1) }{2} } }}{\hat {\bold M} } (x,x') \sqrt{ D(x') e^{-  \frac{ U(x')+U(x'+1) }{2} } }\label{currentMatrixUDsusyq}
\end{eqnarray}
that is the analog of Eq. \ref{susypartnerFPUbiss}.



\section{ Interpreting the supersymmetric partner ${\hat {\bold M} } = - {\bold J} {\bold I}^{\dagger}$ 
as the adjoint ${\mathring {\bold M}}^{\dagger} =- {\mathring {\bold J} }^{\dagger} {\bold I}   $ of  some dual Markov matrix 
${\mathring {\bold M}}=-  {\bold I}^{\dagger} {\mathring {\bold J} }$
}

\label{app_duality}

In this Appendix, we describe how the section \ref{sec_duality} of the main text
can be adapted to the Markov-jump dynamics.

\subsection{ Identification of the dual model }

Let us introduce the shift matrix
\begin{eqnarray}
 {\bold S}(y,x) = \delta_{y,x+1}  
\label{shiftmatrix}
\end{eqnarray}
in order to rewrite the adjoint ${\bold I}^{\dagger} $ of Eq. \ref{IncidenceMatrixTransposed} 
in terms of ${\bold I} $ via the product
\begin{eqnarray}
 {\bold I}^{\dagger} = - {\bold S} {\bold I}
\label{IncidenceMatrixTransposedSm}
\end{eqnarray}
in order to take into account the relations between their matrix elements of Eqs \ref{IncidenceMatrixTransposed}
and 
\ref{IncidenceMatrixTransposed}
\begin{eqnarray}
 {\bold I}^{\dagger} (x'+1,x) = \delta_{x'+1,x}  -  \delta_{x',x} = - {\bold I} (x',x)
\label{IncidenceMatrixTransposedS}
\end{eqnarray}

Then the identification between the supersymmetric partner 
\begin{eqnarray}
{\hat {\bold M} } = - {\bold J} {\bold I}^{\dagger} = {\bold J} {\bold S} {\bold I}
\label{susyJSI}
\end{eqnarray}
and the adjoint ${\mathring {\bold M}}^{\dagger} =- {\mathring {\bold J} }^{\dagger} {\bold I}   $ of  another Markov matrix 
${\mathring {\bold M}}=-  {\bold I}^{\dagger} {\mathring {\bold J} }$ leads to 
\begin{eqnarray}
 {\mathring {\bold J} }^{\dagger} = -  {\bold J} {\bold S} 
\label{Jringdagger}
\end{eqnarray}
that is the analog of Eq. \ref{dualityringcurrent}.

The matrix elements computed from Eq. \ref{currentMatrixUD}
and \ref{shiftmatrix} 
\begin{eqnarray}
 {\mathring {\bold J} }^{\dagger} (x,x') = -  \sum_y {\bold J}(x,y) {\bold S} (y,x')
 =  -   {\bold J}(x,x'+1) 
 = - D(x) e^{-  \frac{ U(x)+U(x+1) }{2} } {\bold I}(x,x'+1) e^{U (x'+1)}
\label{Jringdaggerme}
\end{eqnarray}
yields that the corresponding current matrix elements written using Eq. \ref{IncidenceMatrixTransposedS}
\begin{eqnarray}
 {\mathring {\bold J} } (x',x) && = {\mathring {\bold J} }^{\dagger} (x,x') 
  = - e^{U (x'+1)} {\bold I}^{\dagger}(x'+1,x)   D(x) e^{-  \frac{ U(x)+U(x+1) }{2} } 
  \nonumber \\
  && =   e^{U (x'+1)}  {\bold I} (x',x)   D(x) e^{-  \frac{ U(x)+U(x+1) }{2} } 
\label{Jringme}
\end{eqnarray}
can be identified with the initial form of Eq. \ref{currentMatrixUD} in a dual potential ${\mathring U}(.) $
\begin{eqnarray}
  {\mathring {\bold J} } (x',x) = {\mathring D}(x') e^{-  \frac{ {\mathring U}(x')+{\mathring U}(x'+1) }{2} } {\bold I}(x',x) e^{{\mathring U} (x)}
\label{currentMatrixUDring}
\end{eqnarray}
with the correspondence
\begin{eqnarray}
e^{{\mathring U} (x)} && =  D(x) e^{-  \frac{ U(x)+U(x+1) }{2} } 
\nonumber \\
  e^{U (x'+1)} && = {\mathring D}(x') e^{-  \frac{ {\mathring U}(x')+{\mathring U}(x'+1) }{2} }
\label{dualitydiscrete}
\end{eqnarray}
leading to the dual potential
\begin{eqnarray}
{\mathring U}(x) && = - \frac{U(x)+U(x+1)}{2}  +  \ln D(x) 
 \label{Uring}
\end{eqnarray}
that is the analog of Eq. \ref{Uxring},
and to the dual diffusion coefficient 
\begin{eqnarray}
{\mathring D}(x) && = e^{ U(x+1) + \frac{{\breve U}(x)+{\breve U}(x+1)}{2}} 
= \sqrt{ D(x) D(x+1) e^{  U(x+1) - \frac{U(x)+U(x+2)}{2} }} 
 \label{Dring}
\end{eqnarray}
that replaces the invariance of the diffusion coefficient for the continuous-space case described in the main text.

As discussed after Eq. \ref{siegmund}
in the main text, the transformation of Eqs \ref{Uring}
and \ref{Dring} is related to the notions of Siegmund duality
and to the notion of duality between boundary-driven B.C. and equilibrium B.C.
described in \cite{tailleurMapping,c_boundarydriven}.
It is also related to the notion of duality between random trap models and random barrier models
\cite{jpb_antoine,dyson,alexander,jack1,jack2,jack3}.
Our conclusion is thus that the supersymmetric perspective is very useful to unify various notions
of dualities that have been previously introduced for one-dimensional Markov jump processes.


\subsection{ Interpretation of the duality for the discrete scale function and the discrete speed measure }

Using the correspondence of Eq. \ref{dualitydiscrete},
one obtains that the scale function of Eq. \ref{Sxdiscrete} for the dual model
\begin{eqnarray}
{\mathring s}(y)  =  \sum_{x=0}^{y-1} \frac{ e^{\frac{ {\mathring U}(x)+{\mathring U}(x+1)}{2} }  } { {\mathring D}(x) } 
=  \sum_{x=0}^{y-1} e^{- U(x+1) } = \sum_{x'=1}^{y} e^{- U(x') } = m(y)-m(0)
\label{Sxdiscretering}
\end{eqnarray}
is related to the speed mesure $m(.)$,
while the speed measure of Eq. \ref{mydiscrete} for the dual model
\begin{eqnarray}
{\mathring m}(y)= \sum_{z=0}^y e^{- {\mathring U}(z) } 
= \sum_{z=0}^y  \frac{e^{  \frac{ U(z)+U(z+1) }{2} }}{ D(z)}  =s(y+1)
\label{mydiscretering}
\end{eqnarray}
is related to the scale function $s(.)$.
These relations are the analog of Eq. \ref{speedscalering}.


\section{ Interpreting the supersymmetric partner ${\hat {\bold M} } = - {\bold J} {\bold I}^{\dagger}$ 
as a non-conserved Markov matrix 
 ${\tilde {\bold M}}_{nc}$ 
}

\label{app_nckilling}

In this Appendix, we describe how the section \ref{sec_nckilling} of the main text
can be adapted to the Markov-jump dynamics.

\subsection{ Correspondence between the supersymmetric partner ${\hat {\bold M} } = - {\bold J} {\bold I}^{\dagger}$
and some non-conserved Markov matrix 
 ${\tilde {\bold M}}_{nc}$ involving killing}

Let us introduced the Markov matrix 
 ${\tilde {\bold M}}$ that has the same off-diagonal elements of Eq \ref{hatwoff} 
 as the supersymmetric partner ${\hat {\bold M} } $
 \begin{eqnarray}
{\tilde {\bold M} }(x+1,x) = {\hat {\bold M} } (x+1,x) && = w(x+2,x+1) 
\nonumber \\
{\tilde {\bold M} }(x-1,x) ={\hat {\bold M} } (x-1,x) && = w(x-1,x) 
\label{hatwoffconserved}
\end{eqnarray}
while its diagonal element $ {\tilde {\bold M} }(x,x) $ is the opposite of the total rate out of site $x$ 
as in Eq. \ref{masterdiag} in order to conserve the total probability
\begin{eqnarray}
 {\tilde {\bold M} }(x,x) =   - {\tilde {\bold M} }(x+1,x) - {\tilde {\bold M} } (x-1,x)
 = - w(x+2,x+1) - w(x-1,x) 
\label{masterdiagtilde}
\end{eqnarray}

The difference between this diagonal element
and the diagonal element $ {\hat {\bold M}} (x,x) $ 
of Eq. \ref{hatwdiag} for the supersymmetric partner
can be interpreted as the killing rate
\begin{eqnarray}
{\tilde K} (x) && =  {\tilde {\bold M} }(x,x)- {\hat {\bold M}} (x,x) 
\nonumber \\
&&=  w(x+1,x) + w(x,x+1) - w(x+2,x+1) -w(x-1,x) 
\label{killingtilde}
\end{eqnarray}
that enters the non-conserved version ${\tilde {\bold M} }_{nc}(x,z) $ of the matrix  ${\tilde {\bold M} }$
\begin{eqnarray}
 {\tilde {\bold M} }_{nc}(x,z) \equiv  {\tilde {\bold M} }(x,z) - {\tilde K} (x) \delta_{x,z}
 \label{defmnc}
\end{eqnarray}
that is useful to reinterpret the supersymmetric partner ${\hat {\bold M}} $ as
\begin{eqnarray}
{\hat {\bold M}}= {\tilde {\bold M} }_{nc}
 \label{iddefmnc}
\end{eqnarray}

With these notations, the dynamics of Eq. \ref{jlinkdyn} for the current is governed by the non-conserved 
Markov matrix ${\tilde {\bold M} }_{nc} $ in the bulk that is the analog of the non-conserved Fokker-Planck generator of Eq. \ref{susypartnerFPrew}
\begin{eqnarray}
\partial_t J_t(x) && = \sum_{z=0}^{N-1}   {\hat {\bold M}}_{nc} (x,z) J_t(z) 
\nonumber \\
&& =  {\tilde {\bold M} } (x,x-1) J_t(x-1) 
+  {\tilde {\bold M} } (x,x+1) J_t(x+1) 
+    {\tilde {\bold M} } (x,x) J_t(x) 
 - {\tilde K} (x) J_t(x)
\label{jlinkdynkilling}
\end{eqnarray}
while the vanishing-current boundary conditions $J_t(-1)  =0 =J_t(N) $ of Eq. \ref{vanishJDiscrete}
correspond to absorbing B.C. for the variable $J_t(x)$.


\subsection{ Simplifications for discrete Pearson Markov jump processes with shape-invariance via supersymmetric partnership}

The analog of Pearson diffusions of Eqs \ref{pearsonF} and \ref{pearsonD}
are the Markov jump processes characterized by quadratic transition rates
that should vanish at the boundaries $x_L$ and $x_R$ when they are finite
\begin{eqnarray}
 w(x+1,x) && = a  x^2  + b_+ x + c_+ >0 \text { for $x_L \leq x \leq x_R-1$ \ \ \ and $w(x_R+1,x_R)=0$ if $x_R$ finite} 
\nonumber \\
w(x-1,x) && =  a x^2 +b_- x +c_- >0 \text { for $x_L +1\leq x \leq x_R$ \ \ \ and $w(x_L-1,x_L)=0$ if $x_L$ finite} 
\label{ratesquadratic}
\end{eqnarray}

The off-diagonal elements of Eq. \ref{hatwoffconserved} remain quadratic
\begin{eqnarray}
{\tilde {\bold M} }(x+1,x) && = w(x+2,x+1) = a  (x+1)^2  + b_+ (x+1) + c_+ >0 
 \text { for $x_L \leq x \leq x_R-2$ \ \   and ${\tilde {\bold M} }(x_R,x_R-1)=0$} 
\nonumber \\
{\tilde {\bold M} }(x-1,x) && = w(x-1,x) = a x^2 +b_- x +c_- >0 
\text { for $x_L +1\leq x \leq x_R$ \ \ \ and $w(x_L-1,x_L)=0$ }
\label{newrates}
\end{eqnarray}
where the three parameters $(a,b_-,c_-)$ remain unchanged while the two parameters $(b_+,c_+)$ are changed into
\begin{eqnarray}
{\tilde b}_+ &&= 2a  + b_+ 
\nonumber \\
{\tilde c}_+ && = a+b_+ +c_+
\label{tildebcplus}
\end{eqnarray}
Note that the vanishing ${\tilde {\bold M} }(x_R,x_R-1)=0$ occurs at the shifted position $x=x_R-1$
 since the number of currents living on links is smaller than the number of sites.

The killing rate of Eq. \ref{hatwoffconserved}
\begin{eqnarray}
{\tilde K} (x) && =  w(x+1,x)  - w(x+2,x+1) + w(x,x+1) -w(x-1,x) 
\nonumber \\
&&=  a  x^2  + b_+ x + c_+ - a  (x+1)^2  - b_+ (x+1) - c_+
+ a (x+1)^2 +b_- (x+1) +c_-  - a x^2 -b_- x -c_-
\nonumber \\
&&= b_-  -b_+   
\label{killingpearson}
\end{eqnarray}
is independent of the position $x$ and reduces to the difference $(b_-  -b_+)$ of the two parameters $b_{\pm}$.

The conclusion is thus that if $ {\bold M}^{(a,b_-,c_-,b_+,c_+;x_L,x_R)}$ is the Markov matrix of the Pearson diffusion with the rates
of Eq. \ref{ratesquadratic}, then 
its supersymmetric partner ${\hat {\bold M}}^{(a,b_-,c_-,b_+,c_+;x_L,x_R)}= {\tilde {\bold M} }_{nc} $
is the Pearson diffusion $ {\bold M}^{(a,b_-,c_-,{\tilde b_+},\tilde{c_+};x_L,x_R-1)}$ with the modified rates
of Eq. \ref{newrates} in the presence of the constant killing rate ${\tilde K} = b_-  -b_+ $ of Eq. \ref{killingpearson}
\begin{eqnarray}
{\hat {\bold M}}^{(a,b_-,c_-,b_+,c_+;x_L,x_R)} = {\tilde {\bold M} }_{nc} 
={\bold M}^{(a,b_-,c_-,{\tilde b_+},\tilde{c_+};x_L,x_R-1)}
+ (b_--b_+) {\mathbb 1}
\label{shapeInvarianceDiscrete}
\end{eqnarray}
This shape-invariance property is the analog of Eq. \ref{mappingPearson} of the main text.

As a consequence, as in Eq. \ref{E1Pearson},
the first-excited eigenvalue $E_1$ of the Pearson model of Eq. \ref{ratesquadratic}
is given by the killing rate of Eq. \ref{killingpearson}
\begin{eqnarray}
E_1 = b_--b_+
\label{E1Pearsondiscrete}
\end{eqnarray}
while the corresponding eigenvector $l_1(y)$ is determined by Eq. \ref{spectrallinexpli} with $i_1(x)=1$
and \ref{currentMatrix}
\begin{eqnarray}
 E_1  &&    = l_1(x)  - l_1(x+1)
\nonumber \\
  l_1 (x) &&  =   w(x+1,x)   -  w(x-1,x) =  (b_+ -b_-) x  + (c_ + -c_-)
\label{spectrallinexplin1}
\end{eqnarray}
and thus reduce to a polynomial of degree one.
Via iteration, one can compute all the excited eigenvalues $E_n$ of the initial discrete Pearson model
with their corresponding eigenvectors $l_n$ that reduce to polynomials of degree $n$
(see the reviews \cite{Odake,Sasaki,Sasaki_short,Odake_Long} with different scopes on exactly-solvable birth-death models).

In conclusion, the interpretation of the supersymmetric partner ${\hat {\bold M} } = - {\bold J} {\bold I}^{\dagger}$
as some non-conserved Markov matrix ${\tilde {\bold M}}_{nc}$ involving the killing rate $ {\tilde K }(x)$
is useful to see why the discrete Pearson diffusions are exactly solvable with the shape-invariance property of
that involves a constant killing rate.




\begin{thebibliography}{99}


\bibitem{gardiner}
 C. W. Gardiner, “ Handbook of Stochastic Methods: for Physics, Chemistry and the Natural Sciences” (Springer Series
in Synergetics), Berlin (1985).

\bibitem{vankampen}
N.G. Van Kampen, “Stochastic processes in physics and chemistry”, Elsevier Amsterdam (1992).

\bibitem{risken}
 H. Risken, “The Fokker-Planck equation : methods of solutions and applications”, Springer Verlag Berlin (1989).
 
\bibitem{glauber}
 R.J. Glauber, J. Math. Phys. 4, 294 (1963).
 
  \bibitem{Felderhof}
 B.U. Felderhof, Rev. Math. Phys. 1, 215 (1970); Rev. Math. Phys. 2, 151 (1971).
 
\bibitem{referee}
 C. F. Polnaszek, J. H. Freed,  J. Chem. Phys. 58, 3185 (1973).

\bibitem{siggia}
 E. D. Siggia, Phys. Rev. B 16, 2319 (1977).

\bibitem{kimball}
J. C. Kimball, J. Stat. Phys. 21, 289 (1979).

\bibitem{peschel}
 I. Peschel and V. J. Emery, Z. Phys. B 43, 241 (1981)
 
\bibitem{jpb_antoine}
 J.P. Bouchaud and A. Georges, Phys. Rep. 195, 127 (1990).
 
 \bibitem{pierre}
 C. Monthus and P. Le Doussal, Phys. Rev. E 65 (2002) 66129.
 
\bibitem{texier}
 C. Texier and C. Hagendorf, Europhys. Lett. 86 (2009) 37011.
 
\bibitem{us_eigenvaluemethod}
 C. Monthus and T. Garel, J. Stat. Mech. P12017 (2009).
 
\bibitem{Castelnovo}
 C. Castelnovo, C. Chamon and D. Sherrington, Phys. Rev. B 81, 184303 (2012).
 
 \bibitem{c_lyapunov}
C. Monthus, J. Stat. Mech. (2021) 033303.

\bibitem{us_gyrator}
A. Mazzolo and C. Monthus, Phys. Rev. E 107, 014101 (2023).

\bibitem{us_kemeny}
A. Mazzolo and C. Monthus, J. Stat. Mech. (2023) 063204.

 
\bibitem{c_pearson}
C. Monthus, J. Stat. Mech. (2023) 083204.
 
 \bibitem{c_boundarydriven}
C. Monthus,  J. Stat. Mech. (2023) 063206.



 \bibitem{c_largedevDBsusy}
 C. Monthus, J. Stat. Mech. (2024) 073203.



\bibitem{review_susyquantum}
F. Cooper, A. Khare and U. Sukhatme, Phys. Rep. 251, 267 (1995).

\bibitem{Mielnik_review}
B. Mielnik and O. Rosas-Ortiz, J. Phys. A: Math. Gen. 37 (2004) 10007.


\bibitem{sasaki_reviewSusyQM}
R. Sasaki, The Universe, Vol.2 (2014) No.2 2-32

\bibitem{Review_factorization}
L. Infeld and T. E. Hull, Rev. Mod. Phys. 23, 21 (1951)


 \bibitem{c_susySVD}
C. Monthus, J. Stat. Mech. (2024) 083207


\bibitem{zwanzig}
R. Zwanzig, “Nonequilibrium statistical mechanics” (Oxford University Press, New York, 2001)

\bibitem{ceccato}
A. Ceccato, D. Frezzato,  J. Math. Chem. 57, 1822-1839 (2019)



\bibitem{BookKarlin}
S. Karlin and H. Taylor, 
"A Second Course in Stochastic Processes", Academic Press, New York (1981).

\bibitem{BookYor}
D. Revuz and M. Yor, "Continuous martingales and Brownian motion",
Springer Verlag Berlin Heidelberg (1991)


\bibitem{HandBook}
A. N. Borodin and P. Salminen
"Handbook of Brownian Motion - Facts and Formulae"
Birkhäuser Verlag, Springer Basel (2002)


\bibitem{BookBorodin}
Andrei N. Borodin
"Stochastic Processes", Birkh\"auser
Springer International Publishing 
 Switzerland  (2017)



  \bibitem{mathSusyScaleSpeed}
A. Kuznetsov and M. Yuan, arXiv:2405.11051


\bibitem{tailleurMapping}
J. Tailleur, J. Kurchan and V. Lecomte, J. Phys. A 41, 505001 (2008).


\bibitem{siegmund}
D. Siegmund, Ann. Probab. 4(6): 914 (1976).

\bibitem{Cox1983}
J. T. Cox and U. R\"osler, Stochastic Processes and their Applications 16 (1983) 141


\bibitem{cliff}
P. Clifford and A. Sudbury, Ann. Probab. 13(2): 558-565 (1985).

\bibitem{Dette}
 H. Dette, J. A. Fill, J. Pitman and W. J. Studden, J. Theoret. Probab. 10 (1997) 349. 



\bibitem{siegmund_Intertwining}
T. Huillet and S. Martinez, Vol. 43, No. 2 (2011), pp. 437

\bibitem{Kolo}
V. N. Kolokoltsov, Mathematical Notes 89, 652–660 (2011).

\bibitem{siegmund_pathwise}
A. Sturm, J. M. Swart, Volume 31, pages 932–983, (2018)

\bibitem{Lorek}
P. Lorek, Probability in the Engineering and Informational Sciences, 32(4), 495-521 (2018).

\bibitem{Zhao}
P. Zhao, Acta Mathematica Sinica. English Series; Heidelberg 34, 9: 1460-1472 (2018).

\bibitem{c_DualitySpectral}
C. Monthus, arxiv:2507.11041.

\bibitem{siegmund_runtumble}
M. Gu\'eneau and L. Touzo, J. Phys. A: Math. Theor. 57 225005 (2024)

\bibitem{siegmund_bridge}
M. Gu\'eneau and L. Touzo, J. Stat. Mech. (2024) 083208



\bibitem{levy} 
P. L\'evy, Processus stochastiques et mouvement brownien, Gauthier-Villars, Paris (1948).


\bibitem{Ciesielski-Taylor}
 Z. Ciesielski and S. J. Taylor, Trans. Amer. Math. Soc. 103 (1962) 434

\bibitem{biane}
P. Biane, In Sém. Probabilités Strasbourg, XIX 291, Lecture Notes in Math. 1123. Springer, Berlin, 1985.

\bibitem{toth}
 B. Toth, Ann. Probab. 24 (1996) 1324–1367
 
\bibitem{tourigny} 
A. Comtet and Y. Tourigny, Annales de l’Institut Henri Poincaré - Probabilités et Statistiques 2011, Vol. 47, No. 3, 850



\bibitem{ReviewMohle}
M. M\"ohle, Bernoulli 5(5), (1999) 761


\bibitem{ReviewDuality}
S. Jansen and N. Kurt, Probability Surveys Vol. 11 (2014) 59.

\bibitem{AlgebraicReview}
A. Sturm, J. M. Swart, F. V\"ollering,
Lecture Notes Series, Institute for Mathematical Sciences, National University of Singapore, Genealogies of Interacting Particle Systems, pp. 81-150 (2020)


\bibitem{giardina_particle}
C. Giardina, J. Kurchan, F. Redig, and K. Vafayi, Journal of Statistical Physics, 135(1):25–55, 2009.

\bibitem{giardina_transport}
G. Carinci, C. Giardina, C.o Giberti, F. Redig,
J. Stat. Phys. Volume 152, 657, (2013)

\bibitem{Redig_genetic}
G. Carinci, C. Giardina, C. Giberti, and F. Redig. 
 Stochastic Processes and their Applications, 125(3):941–969, 2015.
 
\bibitem{DualityEigen}
 F. Redig and F. Sau.  In International workshop on
Stochastic Dynamics out of Equilibrium, pages 621. Springer, 2017.


\bibitem{DualityHidden}
R. Frassek, C. Giardina and J. Kurchan, SciPost Phys. 9, 054 (2020)
	


\bibitem{pearson1895}
K. Pearson,  Philos. Trans. R. Soc. Lond. Ser. A 186, 343–414 (1895)

\bibitem{pearson_wong}
E. Wong,  (1964) "The construction of a class of stationary Markoff processes",
in Stochastic processes, in mathematical physics and engineering (ed. R. Bellman), 264–276. American Mathematical Society, Rhode Island.

\bibitem{diaconis}
P. Diaconis and S. Zabell, Statist. Sci. 6(3): 284-302 (1991)

\bibitem{autocorrelation}
B. M. Bibby, I. M. Skovgaard, M. Sorensen, Bernoulli 11(2), 2005, 191–220.

\bibitem{pearson_class}
J. L. Forman and M. Sorensen,
Scandinavian Journal of Statistics, Vol. 35: 438–465, 2008

\bibitem{pearson2012}
G.M. Leonenko, T.N. Phillips,  Journal of Computational and Applied Mathematics 236 (2012) 2853–2868

\bibitem{PearsonHeavyTailed}
F. Avram, N.N. Leonenko, N. Suvak
Markov Processes and  Related Fields, v.19, Issue 2, 249-298 (2013)


\bibitem{pearson2018}
S. Jafarizadeh, IEEE Control Systems Letters, vol. 2, no. 3, pp. 465-470,  2018


\bibitem{TheoExchangeSiegmund}
Th. Assiotis, N. O'Connell, J. Warren,
 Lecture Notes in Mathematics, Sem. Prob.,volume 2252, 301 (2019).

\bibitem{ExchangemsTheo}
Th. Assiotis, Bernoulli 29(2): 1686 (2023).


\bibitem{TheoExactSolv1}
Th. Assiotis, Electron. J. Probab. 23: 1 (2018).


\bibitem{TheoExactSolv2}
Th. Assiotis,  Comm. in Math. Phys. Volume 402, page 2641 (2023).


\bibitem{c_ring}
C. Monthus, J. Stat. Mech. (2019) 023206.




\bibitem{dyson}
F. Dyson, Phys. Rev. 92, 1331 (1953).

\bibitem{alexander}
S. Alexander et al., Rev. Mod. Phys. 53, 175 (1981).

\bibitem{jack1}
R. L. Jack, P. Sollich, J. Phys. A 41, 324001 (2008)

\bibitem{jack2}
R. L. Jack and P. Sollich, J. Stat. Mech. (2009) P11011

\bibitem{jack3}
P. Sollich and R. L. Jack, Progress of Theoretical Physics Supplement No. 184, 200 (2010).


\bibitem{Odake}
S. Odake, R. Sasaki, J. Math. Phys. 49, 053503 (2008)

\bibitem{Sasaki}
R. Sasaki, J. Math. Phys. 50, 103509 (2009).


\bibitem{Sasaki_short}
R. Sasaki, arXiv:1004.4712

\bibitem{Odake_Long}
S. Odake, R. Sasaki, J. Phys. A 44 , 353001 (2011).



\end{thebibliography}
\end{document}